\documentclass{ifacconf}

\usepackage{graphicx}      
\usepackage{natbib}        
\usepackage{amssymb}
\usepackage{amsmath}
\usepackage{algorithm}
\usepackage{algpseudocode}
\usepackage{multirow}
\usepackage{subcaption}
\usepackage{tabularx}

\newcommand{\R}{\mathbb{R}}
\newcommand{\Z}{\mathbb{Z}}
\newcommand{\bma}[1]{\begin{bmatrix}#1\end{bmatrix}}

\newtheorem{assumption}{Assumption}
\newtheorem{corollary}{Corollary}

\begin{document}
\begin{frontmatter}

\title{Capacity Estimation of Lithium-ion Batteries Using Invariance Property in Open Circuit Voltage Relationship\thanksref{footnoteinfo}} 

\thanks[footnoteinfo]{This research has been funded by Volvo AB (Sweden) and NWO (Netherlands) (Grant No. TKI-HTSM 21.0056).}

\author[First]{Yang Wang}
\author[First]{Marta Zagorowska}
\author[First]{Riccardo M.G. Ferrari}

\address[First]{Delft Center for Systems and Control, Delft University of Technology, Mekelweg 5, Delft, 2628 CD, Netherlands (e-mail: y.wang-40@tudelft.nl, m.a.zagorowska@tudelft.nl, r.ferrari@tudelft.nl)}

\begin{abstract}                

Lithium-ion (Li-ion) batteries are ubiquitous in electric vehicles (EVs) as efficient energy storage devices. The reliable operation of Li-ion batteries depends critically on the accurate estimation of battery capacity. However, conventional estimation methods require extensive training datasets from costly battery tests for modeling, and a full cycle of charge and discharge is often needed to estimate the capacity. To overcome these limitations, we propose a novel capacity estimation method that leverages only one cycle of the open-circuit voltage (OCV) test in modeling and allows for estimating the capacity from partial charge or discharge data. Moreover, by applying it with OCV identification algorithms, we can estimate the capacity from dynamic discharge data without requiring dedicated data collection tests. We observed an invariance property in the OCV versus state of charge relationship across aging cycles. Leveraging this invariance, the proposed method estimates the capacity by solving an OCV alignment problem using only the OCV and the discharge capacity data from the battery. Simulation results demonstrate the method's efficacy, achieving a mean absolute relative error of 0.85\% in capacity estimation across 12 samples from 344 aging cycles.

\end{abstract}


\end{frontmatter}

\section{Introduction}
Electric vehicles (EVs) are gaining increasing popularity in modern transportation due to their roles in reducing greenhouse gas emissions, leading to carbon neutralization \citep{meng2018overview,hannan2017review}. Lithium-ion (Li-ion) batteries are renowned for their favorable energy density, energy efficiency, and extended service life, and are widely adopted in EVs as energy storage devices \citep{plett2015battery}. As batteries undergo cycles of charge and discharge, they inevitably experience performance impairment, in which battery capacity declines significantly \citep{peng2025state}. This degradation deteriorates the performance of the EV's range and raises safety concerns. As a result, accurate monitoring of the battery capacity is crucial for reliable vehicle operation \citep{zhang2025study}.

Estimating the capacity of the battery primarily relies on three categories of approaches: direct measurement methods, model-based methods, and data-driven methods \citep{wang2023review}. Direct measurement approaches estimate battery capacity by applying Coulomb counting over a full charge or discharge cycle \citep{navidi2024physics}. This method is straightforward, but it is inconvenient for EVs under operation as the battery is seldom in a fully discharged state, which can permanently damage the battery. Model-based methods enable online estimation of capacity from operational data using established battery models. Common models encompass empirical models and physical models. Empirical model-based approaches collect historical records of the capacity and operating conditions, e.g., the number of cycles, C-rate, and SOC span, and fit an empirical model to the data to relate these characteristics with capacity fade \citep{wan2025degradation,zagorowska2020survey}. Physical model-based methods utilize an electrochemical model (EM) or an equivalent circuit model (ECM) to identify the capacity of the battery. The single particle model (SPM), a commonly used EM, relates microscopic degradation factors, e.g., the mass of active materials and the inventory of lithium-ions, to the battery capacity and simulates the capacity trajectory by modeling the trend of their degradation with the stress factors \citep{el2023physics}. The ECM correlates electrical parameters, e.g., internal resistance, open-circuit voltage (OCV), with battery capacity using data from aging tests. Then, it estimates the capacity during operation by identifying the current values of these parameters from operational data \citep{li2024online}. However, the practical application of EMs is hindered by the difficulty of accessing battery-specific chemical parameters and the requirement for substantial computational power. The ECM needs extensive experiments to establish an accurate correlation between battery parameters and degradation. These factors limit the application of model-based approaches in practical usage. 

Data-driven methods are divided into feature-based methods and end-to-end methods. Feature-based methods establish machine learning models that relate features from the OCV curve, such as the curve in differential voltage analysis (DVA) and incremental capacity analysis (ICA), to the capacity of the battery \citep{navidi2024physics,bloom2005differential}. End-to-end-based methods directly process measurement data, e.g., current, voltage, and temperature from battery operation, and infer from usage records the capacity of the battery using neural network architectures \citep{lu2022battery,lianpo2025capacity}. Although showing promising results, data-driven approaches rely on extensive training data, the acquisition of which is costly and time- and energy-consuming.

A mechanistic model-based capacity estimation was proposed, leveraging the open circuit potential (OCP) of electrodes \citep{dubarry2012synthesize}. The OCV of the full battery is obtained as the difference between the positive and the negative electrode potential. The degradations of active materials of the two electrodes and the loss of lithium-ion inventory are modeled as scaling and translation of the OCP curves, and the degradation of electrode capacity is estimated by fitting alignment parameters of the OCPs to the measured OCV \citep{schmitt2023capacity}. However, this approach requires the data of the half-cell OCP relationships, which rely on invasive experiments that open the battery and build a half-cell from samples of the electrodes. This type of experiment is repeated for each battery configuration, which is costly and induces safety concerns.

In this work, we propose a novel capacity estimation method that leverages a geometric property in the OCV-SOC relationship. The method requires only one cycle of the OCV measurements from the nominal battery, eliminating the need for cumbersome OCV-SOC measurements across all aging cycles and the disassembling of the battery to build half-cell samples. With only a sequence of OCV and discharge capacity data, we estimate the capacity throughout the battery's lifespan with sufficient accuracy. With an OCV estimation algorithm, this approach enables us to estimate the capacity from dynamic discharge data, without using additional capacity diagnostic tests, showing potential for online estimation of capacity from dynamic operational data.

The remainder of this paper is structured as follows. Section \ref{sec: capacity estimation} introduces the proposed capacity estimation method, Section \ref{sec: results} presents the estimation results on real-life batteries, and Section \ref{sec: conclusion} gives concluding remarks of this study.

\section{Battery capacity estimation}\label{sec: capacity estimation}
The estimation of battery capacity relies on a geometric property in the open circuit voltage and state of charge (OCV-SOC) relationship. We first introduce the invariance property observed in the battery OCV relationships, then propose the capacity estimation method utilizing this invariance.

\subsection{State of charge of aged batteries}
The state of charge (SOC) is a critical component closely related to the capacity of the battery. It is defined as the charge stored in the battery in relation to its total capacity. However, the correctness of SOC with respect to the real value of the battery critically depends on the total capacity with which the SOC is computed. For an aged battery with a total capacity of 8 Ah and a nominal capacity of 10 Ah, when the battery is fully discharged, the SOC computed with 10 Ah is 20\% while the actual SOC is 0\%. 

To distinguish the type of SOC, we use notations $a$ and $n$ to denote the aged and the nominal battery. We define $C_i\in\R_{>0}$ as the capacity of the battery and $i\in\{a,n\}$ denotes the type of the battery, e.g., $C_a$ is the aged capacity and $C_n$ is the nominal capacity.

For computing the SOC, we define the discharge capacity $Q_{dc}^i(t)$ as the amount of charge withdrawn from the battery $i$. The value of $Q_{dc}^i(t)$ always starts with zero, independent of the initial SOC when it starts counting. It can be computed via Coulomb counting as:
\begin{align}
    Q_{dc}^{i}(t)=-\frac{1}{3600}\int_{t_0}^ti_b(\tau)d\tau,
\end{align}
where $Q_{dc}^i(t)\in\R$ is the discharge capacity in Ampere-hour (Ah) for battery $i$, $i\in\{a,n\}$ denotes the aged or the nominal battery, $i_b(t)\in\R$ is the load current of the battery (positive for charging), and 3600 is the factor converting Coulombs into Ah. The SOC is defined with $Q_{dc}^i(t)$ and the battery's total capacity $C_j$ as:
\begin{align}\label{eq: SOC definition}
    Z^i_j(t)=Z^i_j(t_0)-\frac{Q_{dc}^i(t)}{C_j}\times 100\%
\end{align}
where $Z^i_j(t)\in[0,1]$ is the SOC of the $i$-th battery computed with the capacity $C_j$ at time $t$, $Z^i_j(t_0)\in[0,1]$ is the initial SOC, and $C_j\in\R_{>0}$ is the total capacity of the battery $j$ in Ah, and $i,j\in\{a,n\}$ denotes the type of the battery. 

When the type of the battery actually being discharged and the type of capacity used to compute the SOC are consistent, i.e., $i=j$, the SOC $Z_j^i(t)$ is defined as calibrated, i.e., $Z_a^a(t)$ and $Z_n^n(t)$ are calibrated, and the capacity is defined as the calibrated capacity, i.e., $C_i$ is calibrated for battery $i$. When $i\neq j$, e.g., $Z_n^a(t)$, the SOC $Z_j^i(t)$ is skewed or uncalibrated, and the capacity is an uncalibrated capacity. Only the calibrated SOC can reflect the real value of the battery currently being operated.

Note that $Z_a^n(t)$ can be defined with the notation, but does not have a physical meaning and is unused in the paper.

The open circuit voltage (OCV) is the terminal voltage when the battery is not connected to external circuits. It is formally measured as the terminal voltage when the battery is in an open circuit and rested for a sufficient amount of time. The value of OCV varies with the SOC, and as we will see in the next subsection, when the SOC is calibrated, the OCV-SOC relationship is invariant across aging cycles. The OCV and SOC are independently collected sequences from the battery. We write by $V_{oc}^i(t)\in\R$, $i\in\{a,n\}$, the OCV of the $i$-th battery, and denote by $V_{oc}(Z(t))$ the OCV-SOC relationship. The query of $V_{oc}(t)$ at $Z(t)$ can be achieved through sample interpolation.

\subsection{Invariance in OCV-SOC relationships}
\begin{figure}[htbp!]
    \centering
    \begin{subfigure}[t]{0.24\textwidth}
        \centering
        \includegraphics[width=\textwidth]{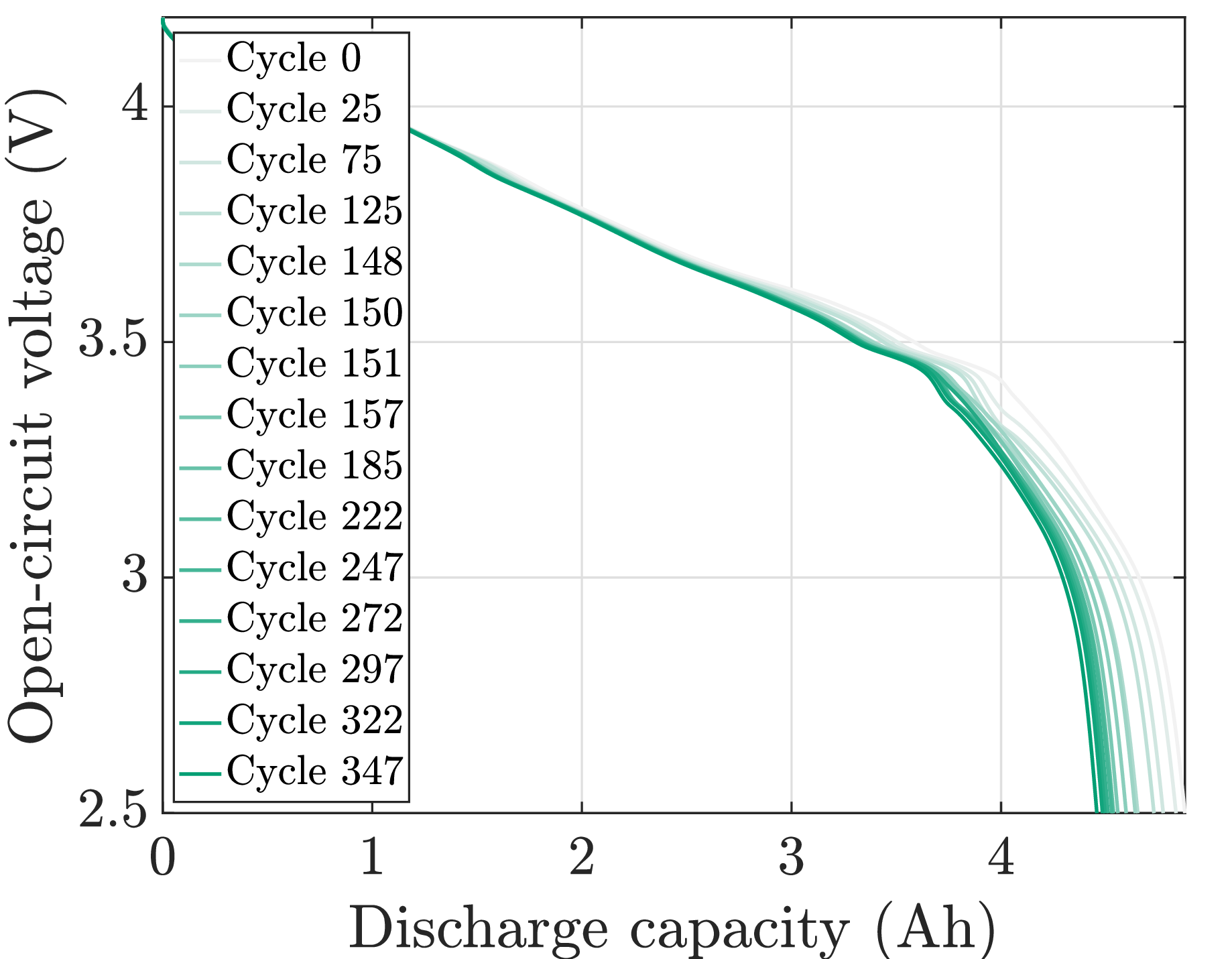}
        \caption{OCV vs discharge capacity of the NMC battery.}
        \label{fig: VQ_curve_stanford}
    \end{subfigure}
    \hfill
    \begin{subfigure}[t]{0.24\textwidth}
        \centering
        \includegraphics[width=\textwidth]{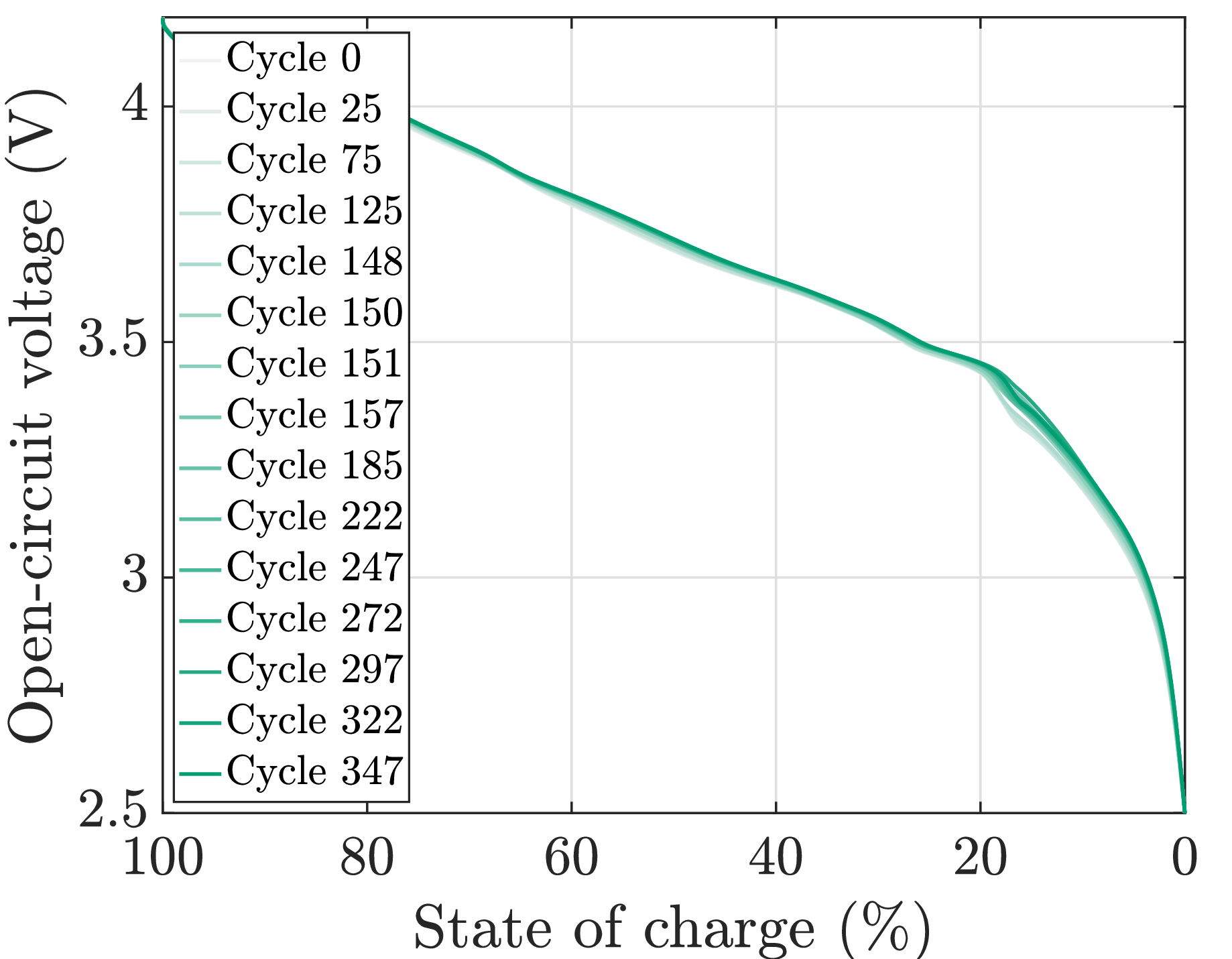}
        \caption{OCV-SOC of the NMC battery.}
        \label{fig: VDoD_curve_stanford}
    \end{subfigure}
    
    \begin{subfigure}[t]{0.24\textwidth}
        \centering
        \includegraphics[width=\textwidth]{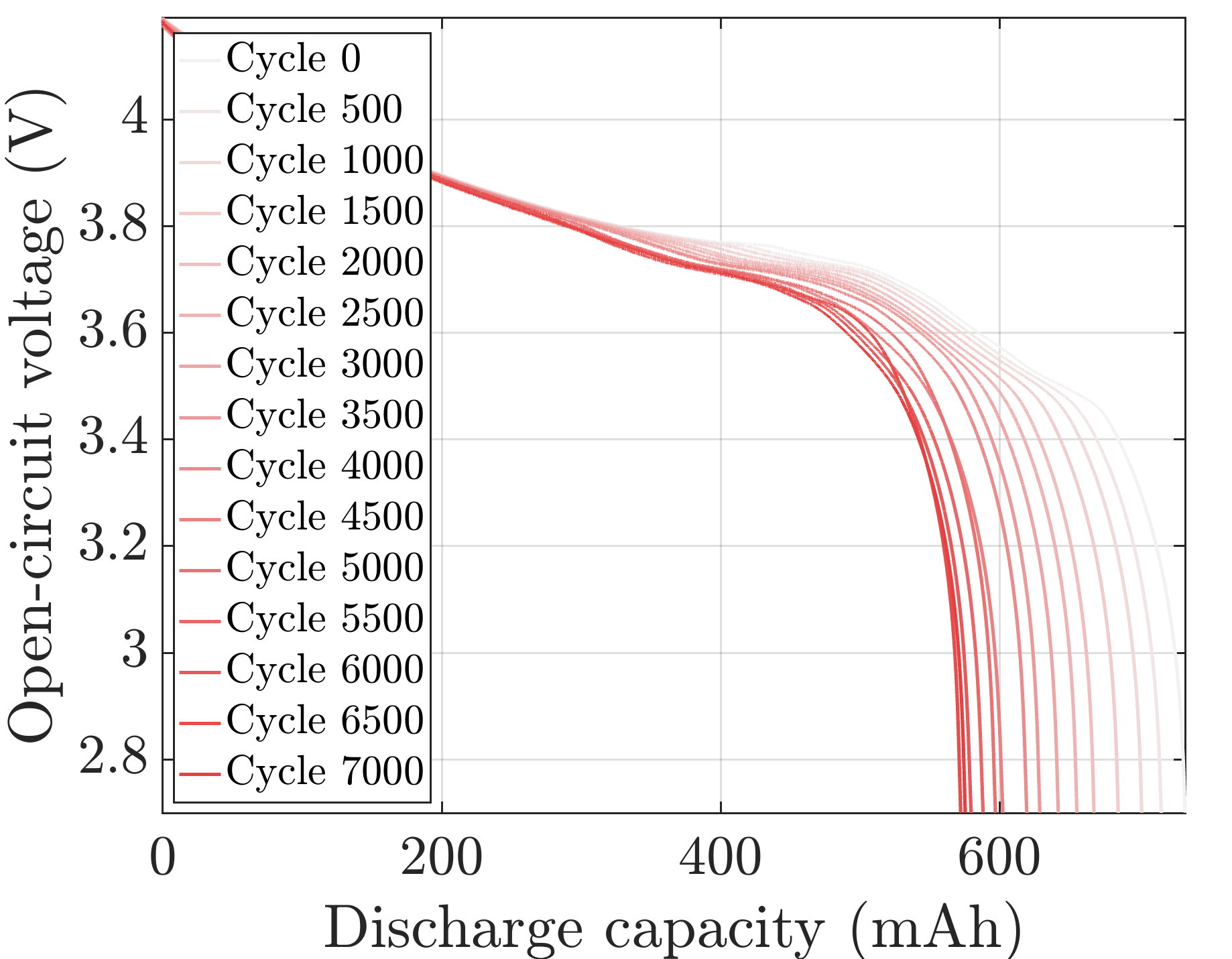}
        \caption{OCV vs discharge capacity of the LMO battery.}
        \label{fig: VQ_curve_oxford}
    \end{subfigure}
    \hfill
    \begin{subfigure}[t]{0.24\textwidth}
        \centering
        \includegraphics[width=\textwidth]{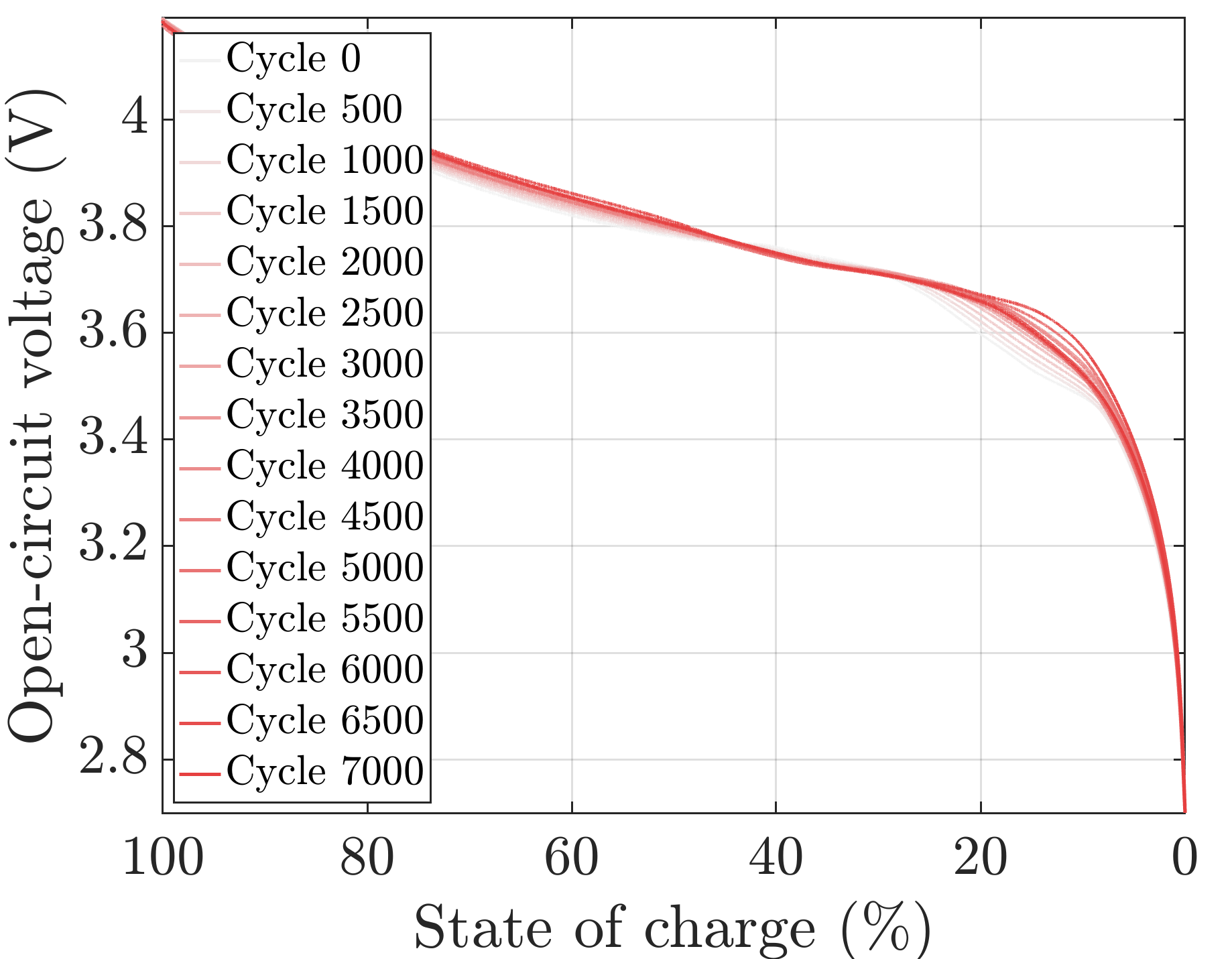}
        \caption{OCV-SOC of the LMO battery.}
        \label{fig: VDoD_curve_oxford}
    \end{subfigure}
    \vskip\baselineskip
    
    \caption{Relationships of OCV versus discharge capacity and OCV-SOC for the NMC \citep{pozzato2022lithium} and the LMO batteries \citep{mcturk2015minimally}.}
    \label{fig:V-Q curve and V-DOD curve of tested batteries}
\end{figure}

The OCV-SOC relationship can reveal the aging status of the battery. When the battery undergoes continuous aging, the total capacity of the battery decreases, and the OCV reaches the lower cut-off voltage faster with less discharge capacity extracted from the battery. This is illustrated in Figures \ref{fig: VQ_curve_stanford} and \ref{fig: VQ_curve_oxford}. However, when we compute the calibrated SOC with the calibrated capacity, we examine in the OCV-SOC relationships that the paths of the relationships are highly invariant across all degrees of aging of the battery. 

This phenomenon is observed in batteries with different types of chemistries from different datasets. For example, Figure \ref{fig:V-Q curve and V-DOD curve of tested batteries} shows the OCV versus discharge capacity relationships and the OCV-SOC relationships of a LiNiMnCoO2 (NMC) battery from the Stanford Accelerated Aging dataset \citep{pozzato2022lithium}. We can see from Figure \ref{fig: VQ_curve_stanford} that the maximum discharge capacity at lower cut-off voltage decreases as the battery undergoes more and more aging cycles, and when the OCV is related to the calibrated SOC computed with the calibrated capacity, we see from Figure \ref{fig: VDoD_curve_stanford} that the OCV-SOC curves align closely with each other across all cycles. The mean of RMSEs of the OCV-SOC relationships compared to the OCV-SOC curve of the nominal battery is 0.0132 V, only 0.31\% of the upper cut-off voltage of the battery. A similar observation can be made from a LiMO2 (LMO) battery using the data from the Oxford Battery Degradation Dataset \citep{mcturk2015minimally} shown in Figures \ref{fig: VQ_curve_oxford} and \ref{fig: VDoD_curve_oxford}. The alignment of OCV-SOC relationships after calculating the calibrated SOC with the calibrated capacity is noticeably close, and the mean RMSE of the OCV curves is 0.0274 V, 0.65\% of the battery's upper cut-off voltage.

From this observation, we develop a novel capacity estimation method using the invariance property in the OCV-SOC relationship. To formulate the estimation method, we make a formal assumption about the OCV-SOC relationship in the following.
\begin{assumption}\label{ass: OCV-SOC invariance}
    The open circuit voltage versus state of charge relationship is invariant under aging when the state of charge is calculated with the battery's calibrated capacity.
\end{assumption}
The shape-invariance assumption is proposed similarly in the half-cell OCP model for capacity estimation. The principle lies in the hypothesis that the thermodynamic property depends only on the charging degree of the cell and is a structural property that would not change with battery age \citep{schmitt2023capacity}.

From Assumption \ref{ass: OCV-SOC invariance}, we can derive the following corollary.
\begin{corollary}\label{colr: OCV-SOC invariance}
    The open circuit voltages are equal for the same level of calibrated SOC computed with the calibrated capacity of the battery.
\end{corollary}
This property is the essence of the capacity estimation method proposed in the next subsection. In addition to the OCV relationship, we make the following assumption on the functional property of the OCV-SOC relationship.
\begin{assumption}\label{ass: bijective OCV}
    The OCV-SOC relationship is bijective at every aging state of the battery.
\end{assumption}
This assumption ensures that every SOC maps to a unique OCV and every OCV has a unique SOC to map, and this gives the uniqueness of the solution in capacity estimation in the next subsection.

\subsection{Capacity estimation based on OCV invariance}
We use the invariance property in the OCV-SOC relationship to estimate the battery capacity. From Corollary \ref{colr: OCV-SOC invariance}, we have that when the calibrated SOCs of the aged and the nominal battery are equal, i.e., $Z_a^a(t)=Z_n^n(t)$, the OCVs of the battery at two aging states equal each other, i.e., $V_{oc}^a(Z_a^a(t))\equiv V_{oc}^n(Z_n^n(t))$, and we have: 
\begin{align}\label{eq: OCV invariance}
    V_{oc}^a(Z_a^a(t))=V_{oc}^n(Z_a^a(t)).
\end{align}
When the OCV $V_{oc}^a(t)$ of the aged battery is obtained, from the nominal OCV-SOC relationship $V_{oc}^n(Z_n^n(t))$, we can find the calibrated SOC $Z_a^a(t)$ by looking for the SOC $Z_n^n(t)$ that gives the same value $V_{oc}^n(Z_n^n(t))$ as the aged OCV $V_{oc}^a(t)$.

As we are given only the aged discharge capacity $Q_{dc}^a(t)$ and the OCV $V_{oc}^a(t)$, we can define the SOC $Z_a^a(t)$ with an unknown calibrated capacity $C_a$ and an initial SOC $Z_a^a(t_0)$ and solve for the capacity $C_a$ via an optimization problem.

Let $C_a\in\R$ be the capacity of the aged battery, $Z_a^a(t_0)\in[0,1]$ be the initial SOC of $Z_a^a(t)$. The calibrated SOC $Z_a^a(t)$ of the aged battery can be expressed in terms of $Q_{dc}^a(t)$ as:
\begin{align}\label{eq: SOC aa}
    Z_a^a(t)=Z_a^a(t_0)-\frac{Q_{dc}^a(t)}{C_a}\times100\%.
\end{align}

By letting the OCV of the nominal OCV-SOC relationship $V_{oc}^n(Z_n^n(t))$ evaluated at the calibrated SOC $Z_a^a(t)$ equal the OCV $V_{oc}^a(t)$ of the aged battery, we can solve for the calibrated capacity $C_a$ and the initial SOC $Z_a^a(t_0)$ by solving the following optimization problem:
\begin{align}\label{eq: capacity estimation problem}
    \min_{C_a,Z_a^a(t_0)}\ \|V_{oc,m}^a-V_{oc,m}^n(Z_{a,m}^a)\|_F^2
\end{align}
where $\|\cdot\|$ is the Frobenius norm, $V_{oc,m}^a\in\R^{m+1}$ is the vector of OCVs of the aged battery $V_{oc}^a(t)$, and $V_{oc,m}^n(Z_{a,m}^a)\in\R^{m+1}$ is the vector of OCVs of the OCV-SOC relationship $V_{oc}^n(Z_n^n(t))$ evaluated at $Z_{a,m}^a$:
\begin{align}
    V_{oc,m}^{a}=\bma{V_{oc}^a(t_0)\\
    V_{oc}^a(t_1)\\
    \vdots\\
    V_{oc}^a(t_m)},\ 
    V_{oc,m}^{n}(Z_{a,m}^a)=\bma{V_{oc}^n(Z_{a}^a(t_0))\\
    V_{oc}^n(Z_{a}^a(t_1))\\
    \vdots\\
    V_{oc}^n(Z_{a}^a(t_m))}.
\end{align}
$Z_{a,m}^a\in\R^{m+1}$ is the vector of SOCs written by discharge capacity $Q_{dc}^a(t)$ and initial value $Z_a^a(t_0)$ with \eqref{eq: SOC aa}:
\begin{align}
    Z_{oc,m}^{n}=Z_a^a(t_0)\mathbf{1}-\frac{1}{C_a}\bma{Q_{dc}^a(t_0)\\
    Q_{dc}^a(t_1)\\
    \vdots\\
    Q_{dc}^a(t_m)},
\end{align}
where $\mathbf{1}\in\R^{m+1}$ is a vector of 1, $t_k:=kT,k\in\Z_1^m$ is the $k$-th time instant, and $T\in\R_{>0}$ is the sampling time. The evaluation $V_{oc,m}^n(Z_{a,m}^a)$ of the OCV-SOC relationship $V_{oc}^n(Z_{a}^n(t))$, composed of two sequences $V_{oc}^n(t)$ and $Z_n^n(t)$, at $Z_{a,m}^a$ can be achieved via linear interpolation.

\begin{figure}[htbp!]
    \centering
    \begin{subfigure}[t]{0.24\textwidth}
        \centering
        \includegraphics[width=\textwidth]{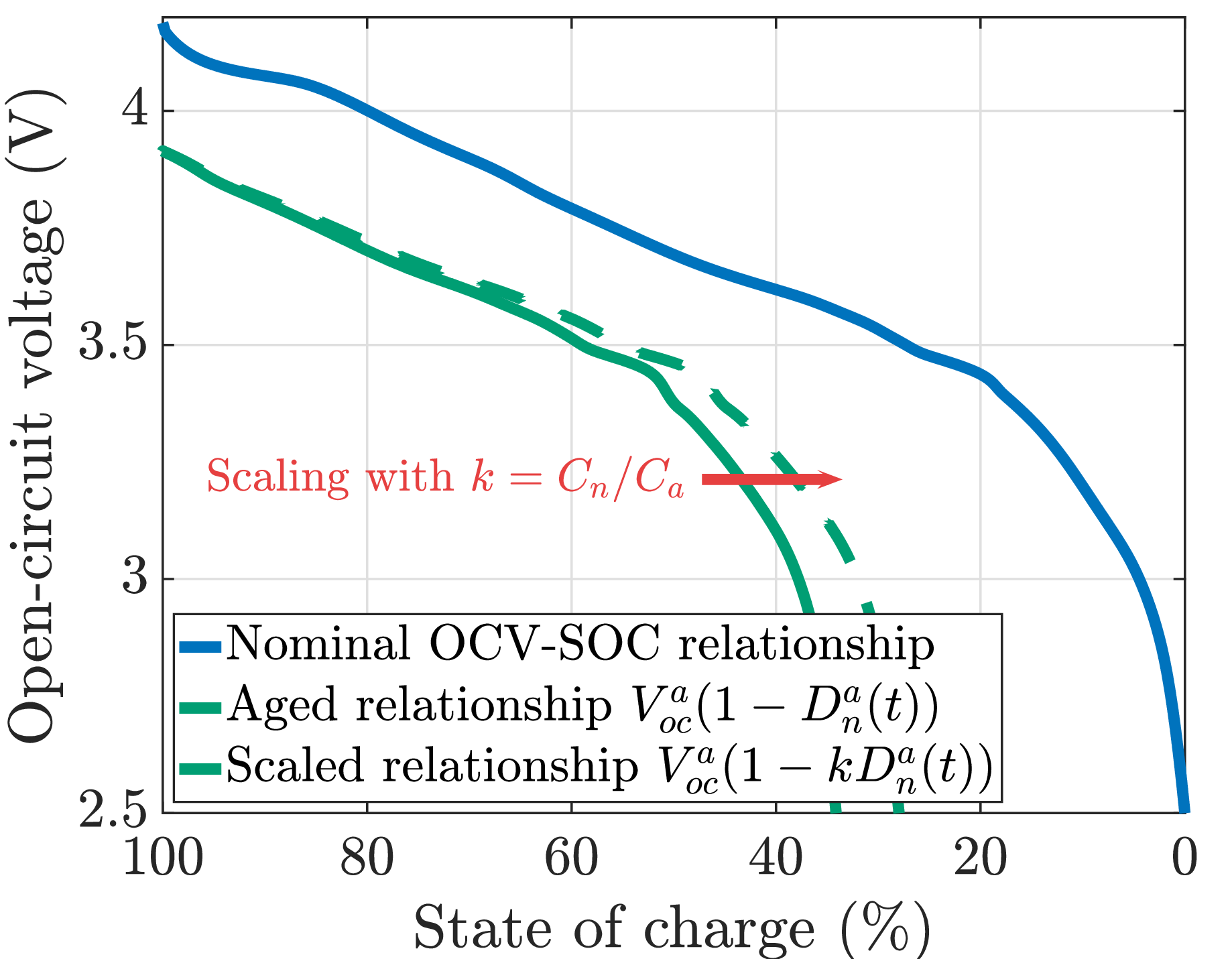}
        \caption{Scaling}
        \label{fig: translation_illustration}
    \end{subfigure}
    \hfill
    \begin{subfigure}[t]{0.24\textwidth}
        \centering
        \includegraphics[width=\textwidth]{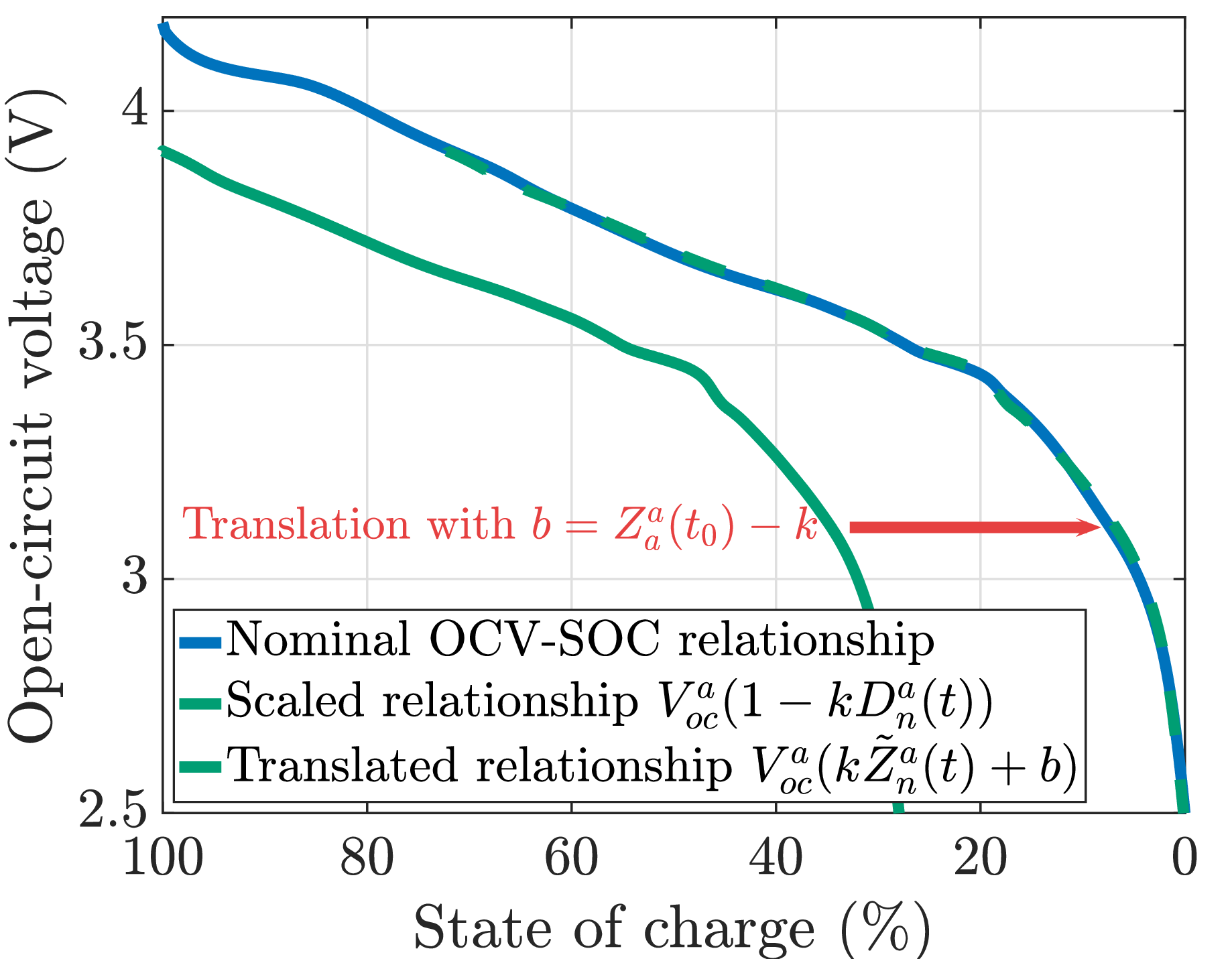}
        \caption{Translation}
        \label{fig: scaling_illustration}
    \end{subfigure}
    \caption{Scaling and translation of the uncalibrated OCV-SOC relationship for capacity estimation.}
    \label{fig: transformation illustration}
\end{figure}

To validate the fulfillment of invariance with the obtained capacity $C_a$ and initial value $Z_a^a(t_0)$, we derive the calibrated OCV-SOC relationship $V_{oc}^a(Z_a^a(t))$ from the discharge capacity $Q_{dc}^a(t)$ and the OCV $V_{oc}^a(t)$, then compare it with the nominal relationship. We define an uncalibrated 1-initialized SOC as $\tilde Z_n^a(t):=1-D_n^a(t)$, where $D_n^a(t):=Q_{dc}^a(t)/C_n$. The calibrated relationship $V_{oc}^a(Z_a^a(t))$ can be obtained from the uncalibrated relationship $V_{oc}^a(\tilde Z_n^a(t))$ by scaling and translating $\tilde Z_n^a(t)$ as:
\begin{align}
    Z_a^a(t)=k\tilde Z_n^a(t)+b,
\end{align}
where $k:=C_n/C_a\in\R$ is the scaling factor and $b:=Z_a^a(t_0)-k\in\R$ is the translation factor of the transformation. This transformation is illustrated by an example shown in Figure \ref{fig: transformation illustration}. The transformed OCV-SOC relationship $V_{oc}^a(Z_a^a(t))$ will align with the nominal relationship $V_{oc}^n(Z_n^n(t))$ satisfying the invariance property.

The developed method utilizes the nominal OCV-SOC relationship $V_{oc}^n(Z_n^n(t))$ and the data of the aged OCV $V_{oc}^a(t)$ and the discharge capacity $Q_{oc}^a(t)$. The OCV value $V_{oc}^a(t)$ can be obtained from dynamic discharge data with an OCV identification algorithm from our previous work \citep{wang2025directcontinuoustimelpvidentification,wang2025continuoustimeidentificationocvreconstruction}. 

The advantages of the developed method are threefold. First, this method requires only one sequence of the OCV-SOC relationship from the nominal battery, without using intensive battery tests. Second, the developed method allows for estimating the capacity from partial discharge data, eliminating the need to fully charge or discharge the battery. Third, by applying it with an OCV identification algorithm, we can estimate the capacity from dynamic discharge data without using additional battery experiments during maintenance.

In the following section, we validate the capacity estimation method using real-life battery data from both OCV tests and dynamic discharge tests.

\section{Simulation Experiments}\label{sec: results}
We evaluate the effectiveness of the developed method for capacity estimation on the Stanford Accelerated Aging dataset \citep{pozzato2022lithium}. The dataset tested a collection of ten battery cells; we selected cell W5 for the validation of this paper. The specification of the tested cell is shown in Table \ref{tab: battery specifications}.

\begin{table}[!htbp]
\centering
\caption{Specifications of the tested battery.}
\begin{tabular}{ll}
\hline
Specification         & Value                           \\ \hline
Battery chemistry     & LiNiMnCoO2/Graphite and silicon \\
Nominal voltage       & 3.63V                            \\
Nominal capacity      & 4.85Ah                         \\
Higher cutoff voltage & 4.2V                            \\
Lower cutoff voltage  & 2.5V                            \\ \hline
\end{tabular}
\label{tab: battery specifications}
\end{table}

The tested cell experienced a total of 344 aging cycles, each of which was discharged with an Urban Dynamometer Driving Schedule (UDDS) profile that emulates real-life driving conditions. The tested battery was discharged from 80\% to 20\% SOC and was charged with a constant current and constant voltage profile. After every few cycles, an offline low-current OCV test was performed to measure the OCV and the current capacity of the battery.

\subsection{Capacity estimation from OCV test data}
We first validate the capacity estimation method using the OCV test data in the accelerated aging dataset. We collected the OCV and the discharge capacity after the 1st, 159th, 194th, 269th, and 344th cycles of the battery as the validation cycles used to test the developed approach. We selected the OCV and the SOC of the first cycle as the nominal OCV-SOC relationship for validation of capacity estimation. The calibrated capacity $C_a$ and the initial SOC $Z_a^a(t_0)$ are estimated by solving \eqref{eq: capacity estimation problem}. The estimated capacities and the absolute relative errors (ARE) of each validation cycle are tabulated in Table \ref{tab: capacity estimation error OCV test}. We see from the table that the estimated capacities exhibit AREs of less than 0.5\% for all 4 validation cycles.

\begin{table}[htbp!]
\centering
\caption{Estimated capacities (Ah) and absolute relative errors (\%) of capacity estimation of the aged battery with OCV test data.}
\begin{tabular}{llll}
\hline
Aging cycle & \begin{tabular}[c]{@{}l@{}}Estimated \\ capacity (Ah)\end{tabular} & \begin{tabular}[c]{@{}l@{}}Actual \\ capacity (Ah)\end{tabular} & ARE (\%) \\ \hline
159 & 4.6280 & 4.6285 & 0.0102 \\
194 & 4.5568 & 4.5466 & 0.2241 \\
269 & 4.5166 & 4.5007 & 0.3538 \\
344 & 4.4706 & 4.4505 & 0.4501 \\ \hline
\end{tabular}\label{tab: capacity estimation error OCV test}
\end{table}

\begin{figure}[htbp!]
    \centering
    \begin{subfigure}[t]{0.24\textwidth}
        \centering
        \includegraphics[width=\textwidth]{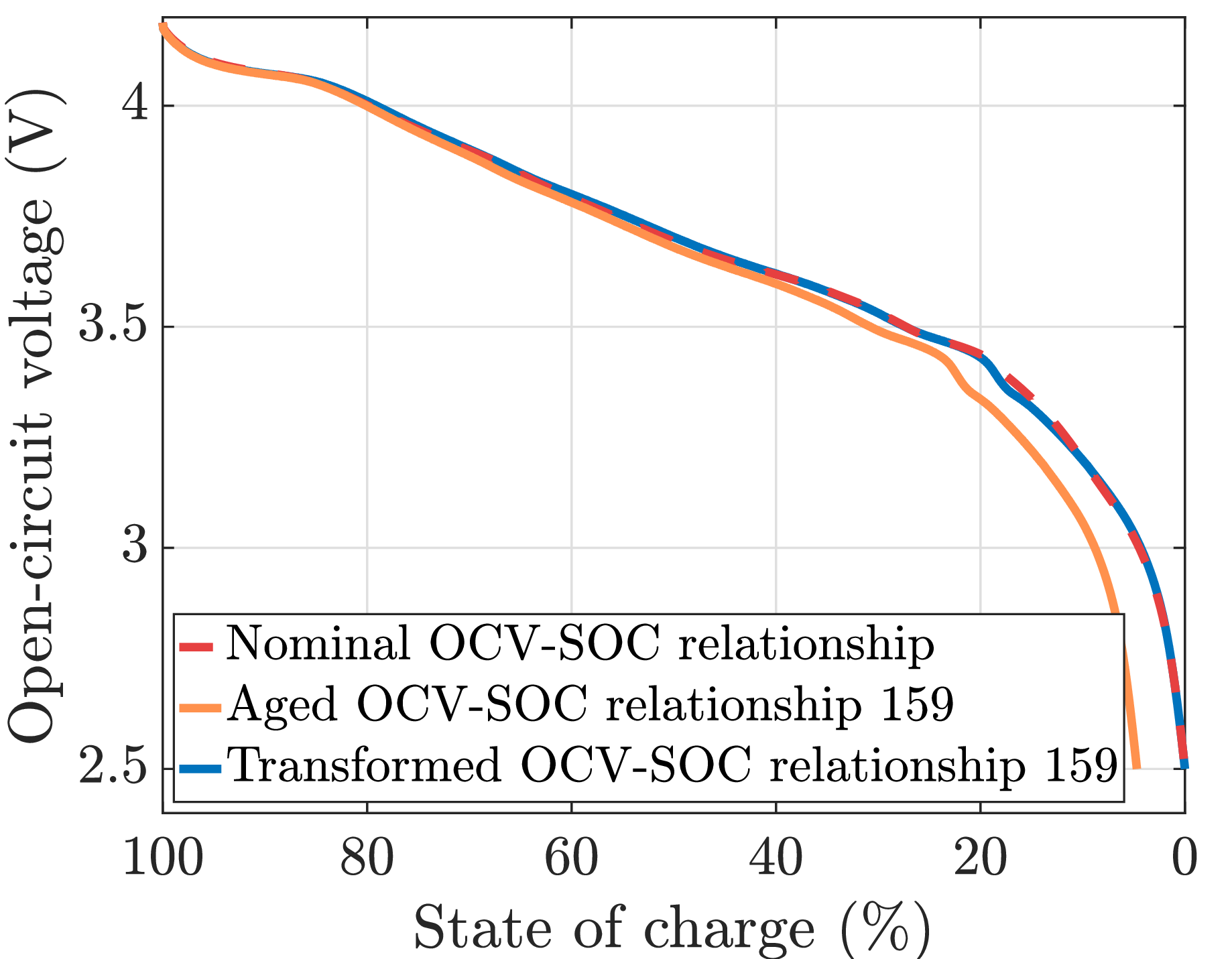}
        \caption{Cycle 159}
        \label{fig: cycle159}
    \end{subfigure}
    \hfill
    \begin{subfigure}[t]{0.24\textwidth}
        \centering
        \includegraphics[width=\textwidth]{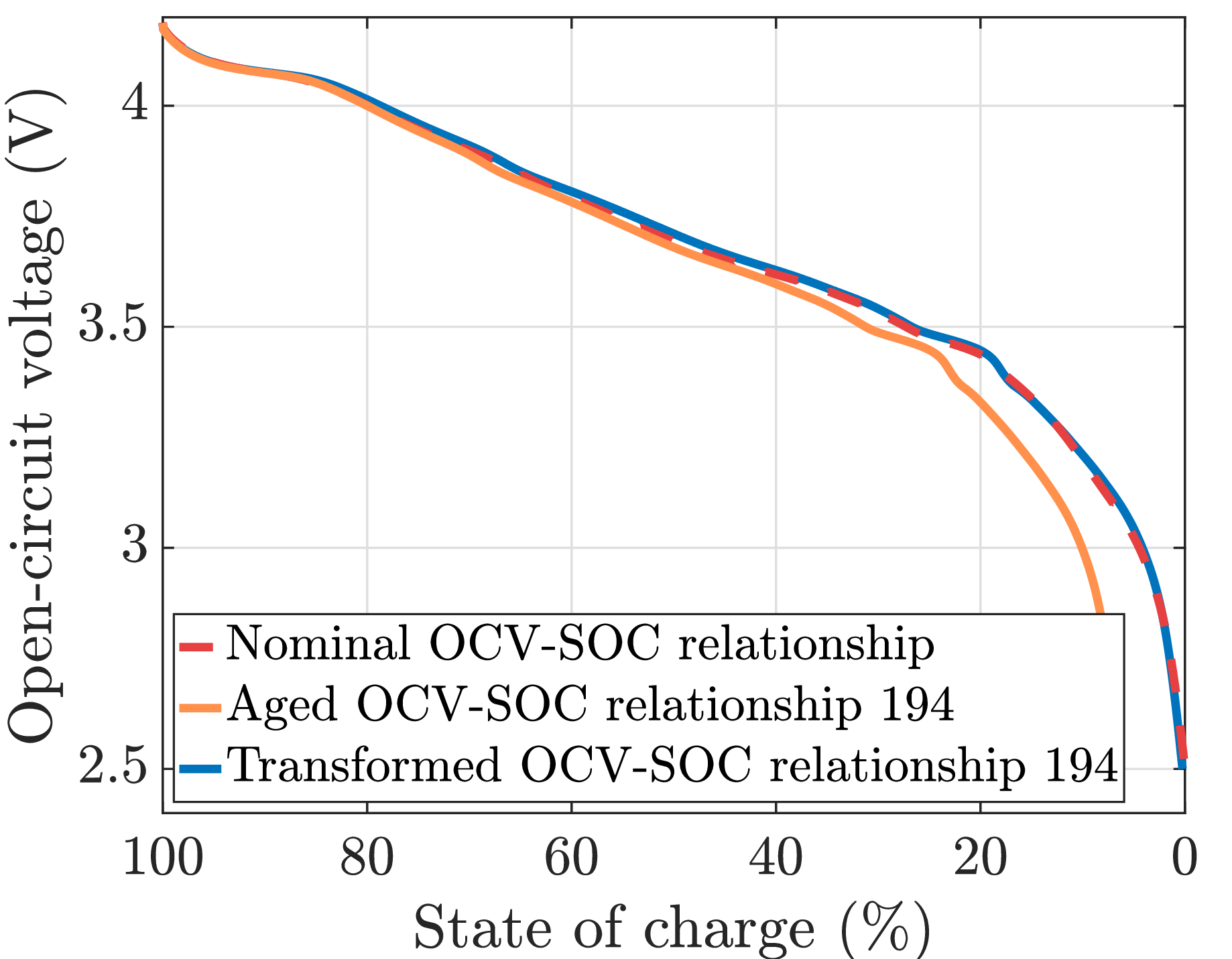}
        \caption{Cycle 194}
        \label{fig: cycle194}
    \end{subfigure}
    \vskip\baselineskip
    
    \begin{subfigure}[t]{0.24\textwidth}
        \centering
        \includegraphics[width=\textwidth]{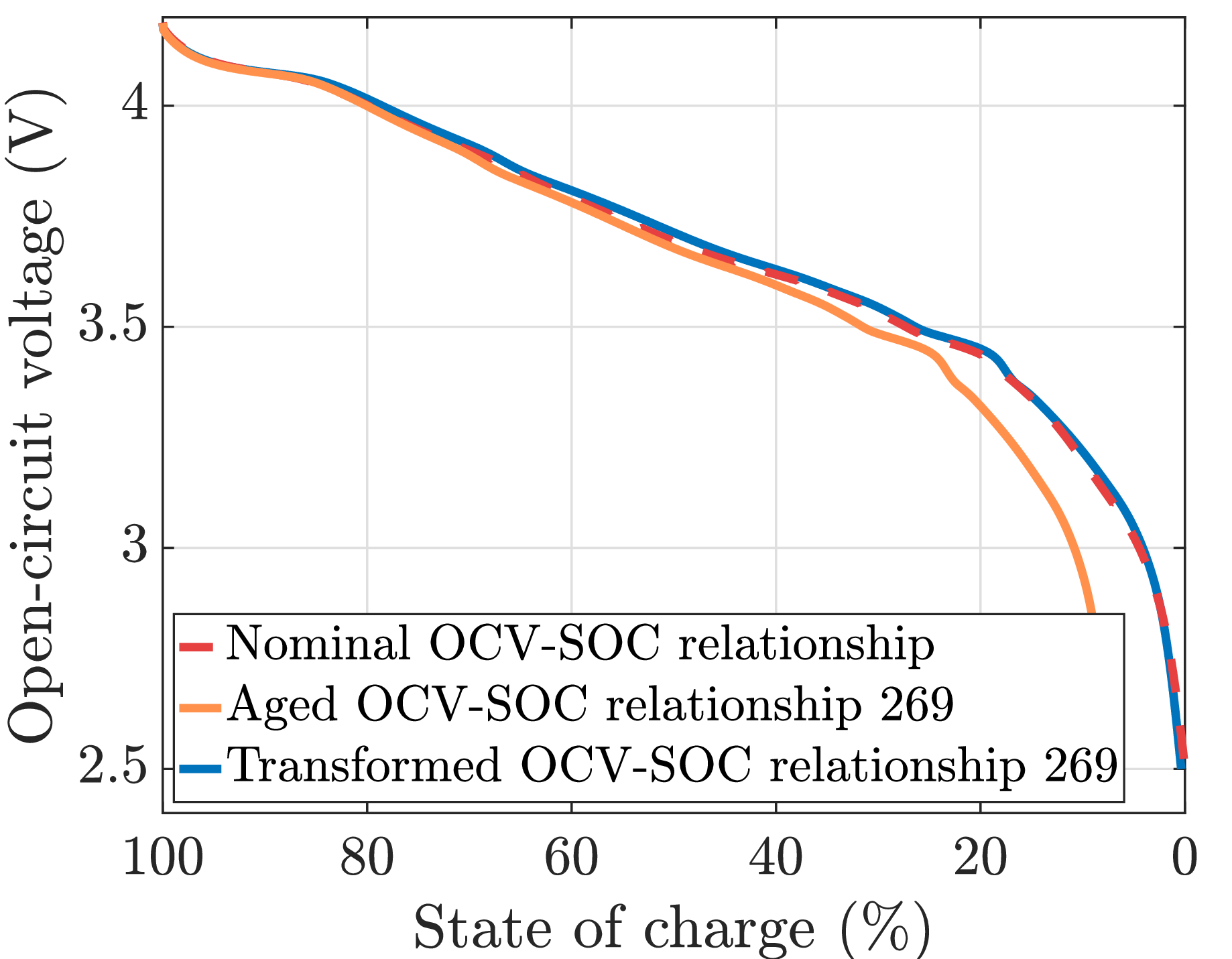}
        \caption{Cycle 269}
        \label{fig: cycle269}
    \end{subfigure}
    \hfill
    \begin{subfigure}[t]{0.24\textwidth}
        \centering
        \includegraphics[width=\textwidth]{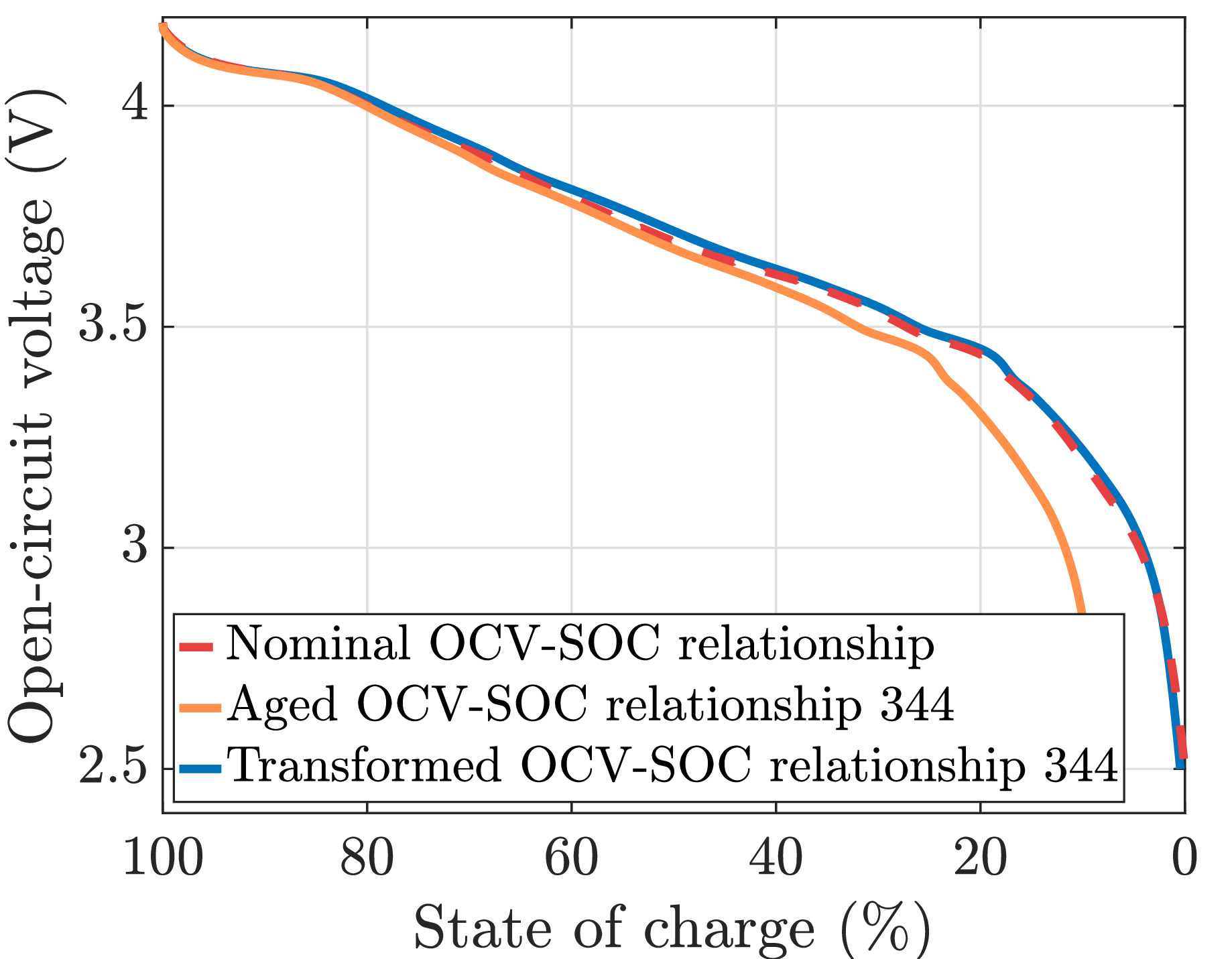}
        \caption{Cycle 344}
        \label{fig: cycle344}
    \end{subfigure}
    \vskip\baselineskip
    
    \caption{OCV-SOC transformations of capacity estimation of validation cycles using OCV test data.}
    \label{fig: capacity estimation OCV test}
\end{figure}

The OCV-SOC transformations of the 4 cycles are shown in Figure \ref{fig: capacity estimation OCV test}. From the figure, we see that the calibrated OCV-SOC relationships with the estimated capacities and initial SOCs satisfy the invariance property.

We validate the effectiveness of the developed method on partial OCV data. The data from cycles 159 and 194 are tested. For each cycle, the first, middle, and last 33\% of the OCV data are used to estimate the capacity of the battery. The AREs of each segment of data of the two cycles, and the estimated capacities of the battery are shown in Table \ref{tab: capacity estimation error OCV test partial}. The table demonstrates an average of less than 1.5\% relative error of capacity estimation of the aged battery using partial OCV data. The average estimated capacities from three parts of the OCV data are 4.69 Ah and 4.61 Ah for the 159th and 194th cycles, respectively. The actual capacities for the two cycles are 4.63 Ah and 4.55 Ah, showing close agreement.

\begin{table}[htbp!]
\centering
\caption{AREs (\%) of capacity estimation using partial OCV data for the 159th and 194th cycles of the aging test.}
\begin{tabular}{lllll}
\hline
Cycle   & Beginning & Middle & End    & Average   \\ \hline
Cycle 159 & 2.6949    & 1.0627 & 2.4023 & 1.3448 \\
Cycle 194 & 5.5931    & 1.8138 & 2.9077 & 1.4997 \\ \hline
\end{tabular}\label{tab: capacity estimation error OCV test partial}
\end{table}

\begin{figure}[htbp!]
    \centering
    \begin{subfigure}[t]{0.24\textwidth}
        \centering
        \includegraphics[width=\textwidth]{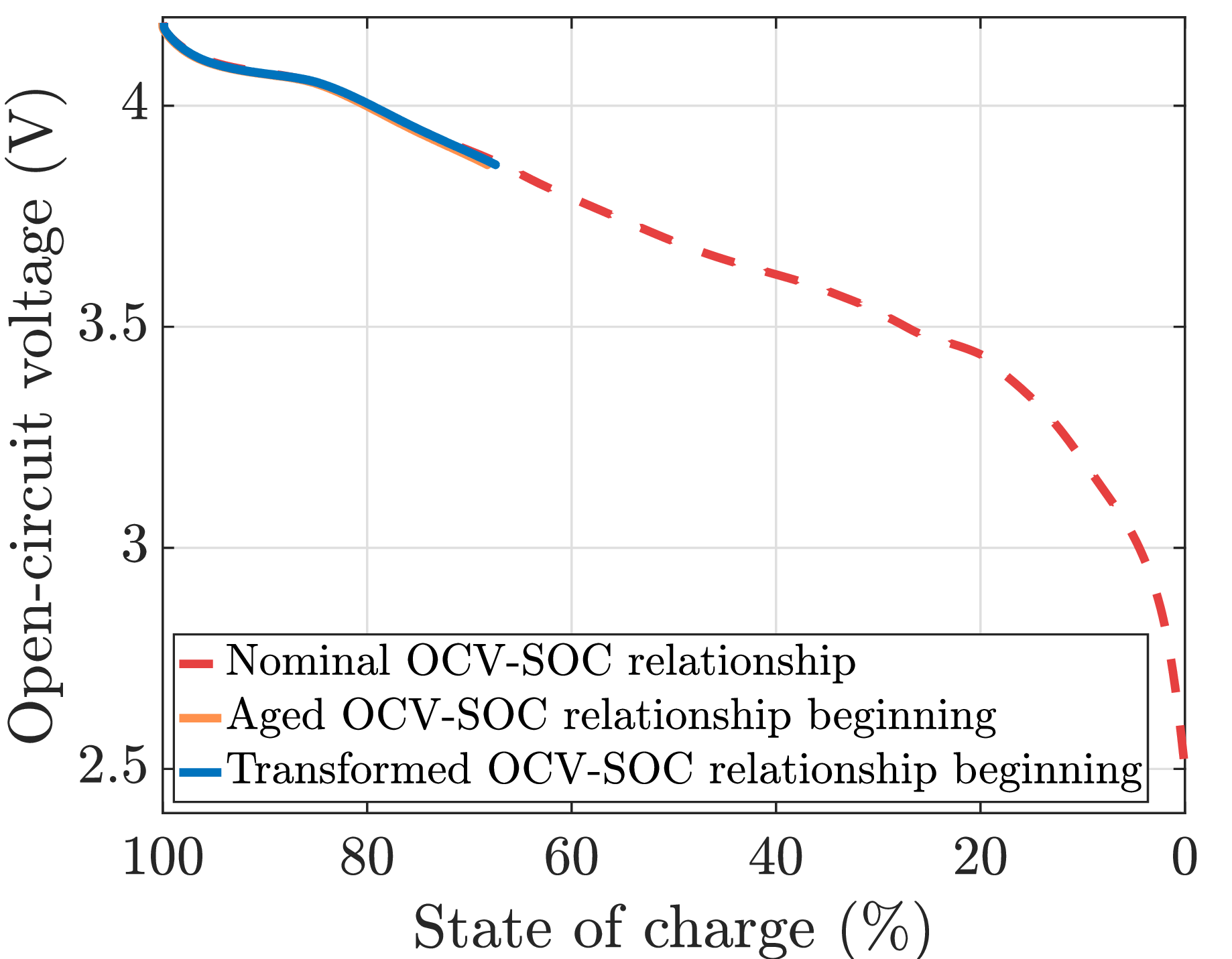}
        \caption{Cycle 159 beginning}
        \label{fig:cycle159_beginning}
    \end{subfigure}
    \hfill
    \begin{subfigure}[t]{0.24\textwidth}
        \centering
        \includegraphics[width=\textwidth]{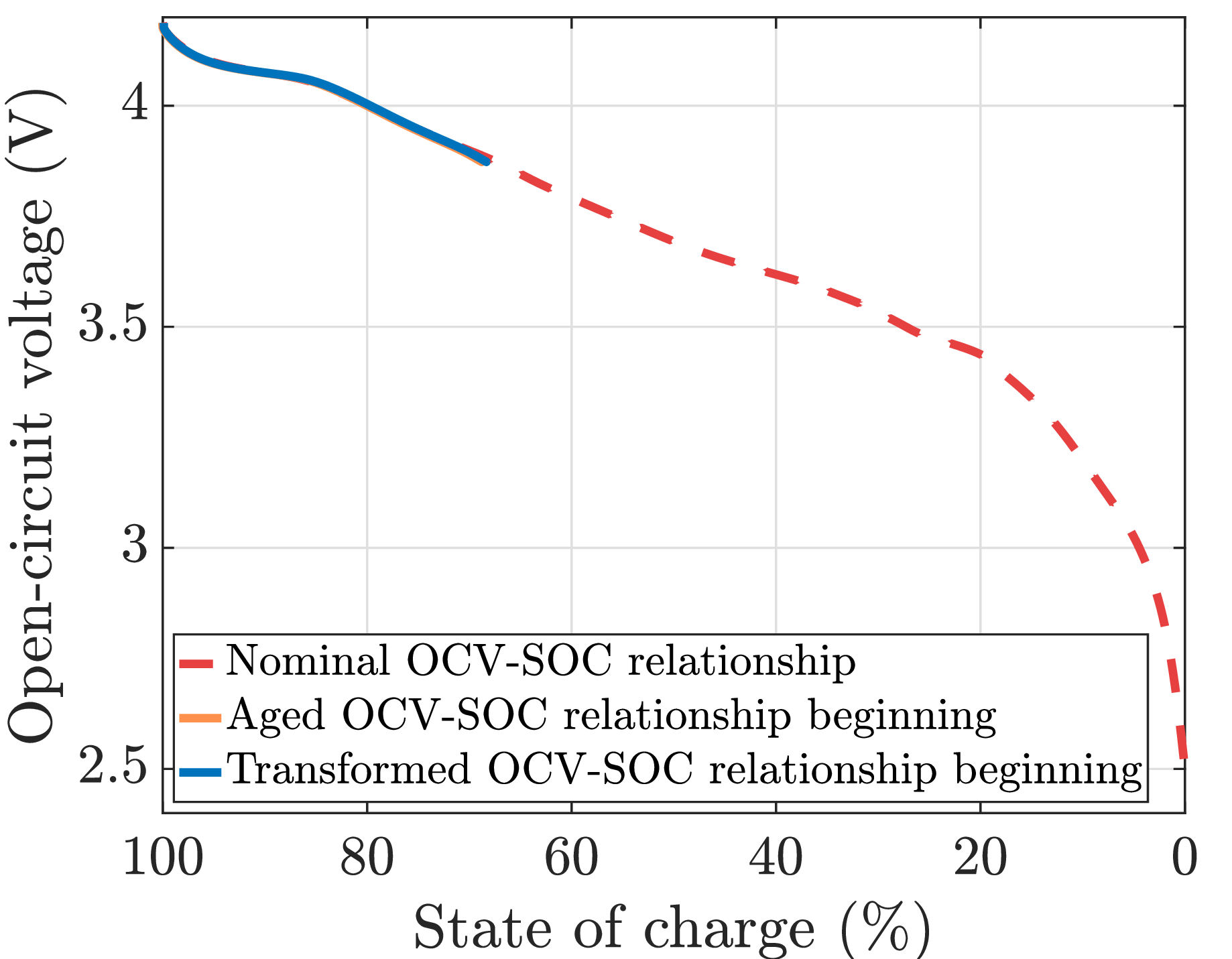}
        \caption{Cycle 194 beginning}
        \label{fig:cycle194_beginning}
    \end{subfigure}

    \vspace{\baselineskip}

    \begin{subfigure}[t]{0.24\textwidth}
        \centering
        \includegraphics[width=\textwidth]{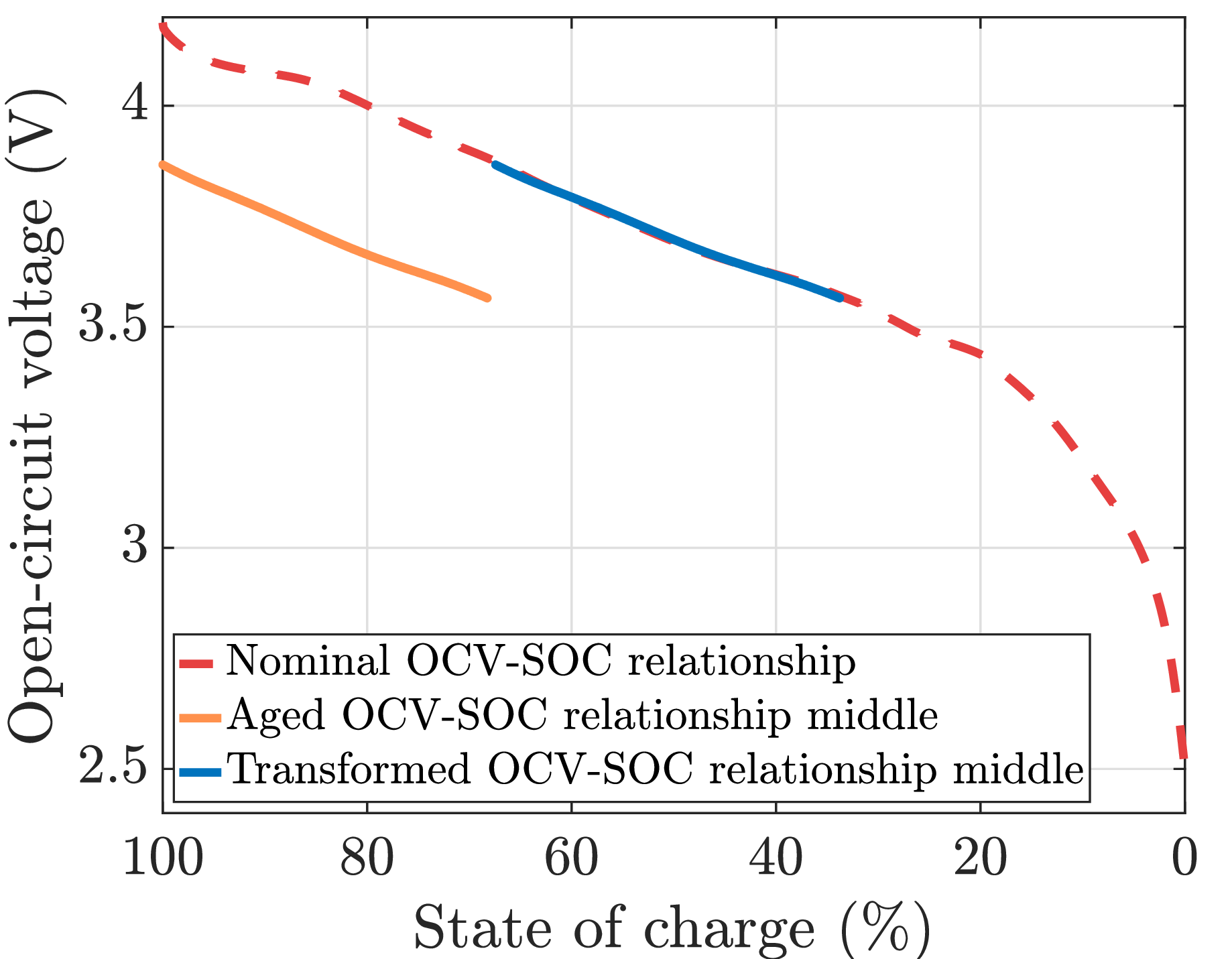}
        \caption{Cycle 159 middle}
        \label{fig:cycle159_middle}
    \end{subfigure}
    \hfill
    \begin{subfigure}[t]{0.24\textwidth}
        \centering
        \includegraphics[width=\textwidth]{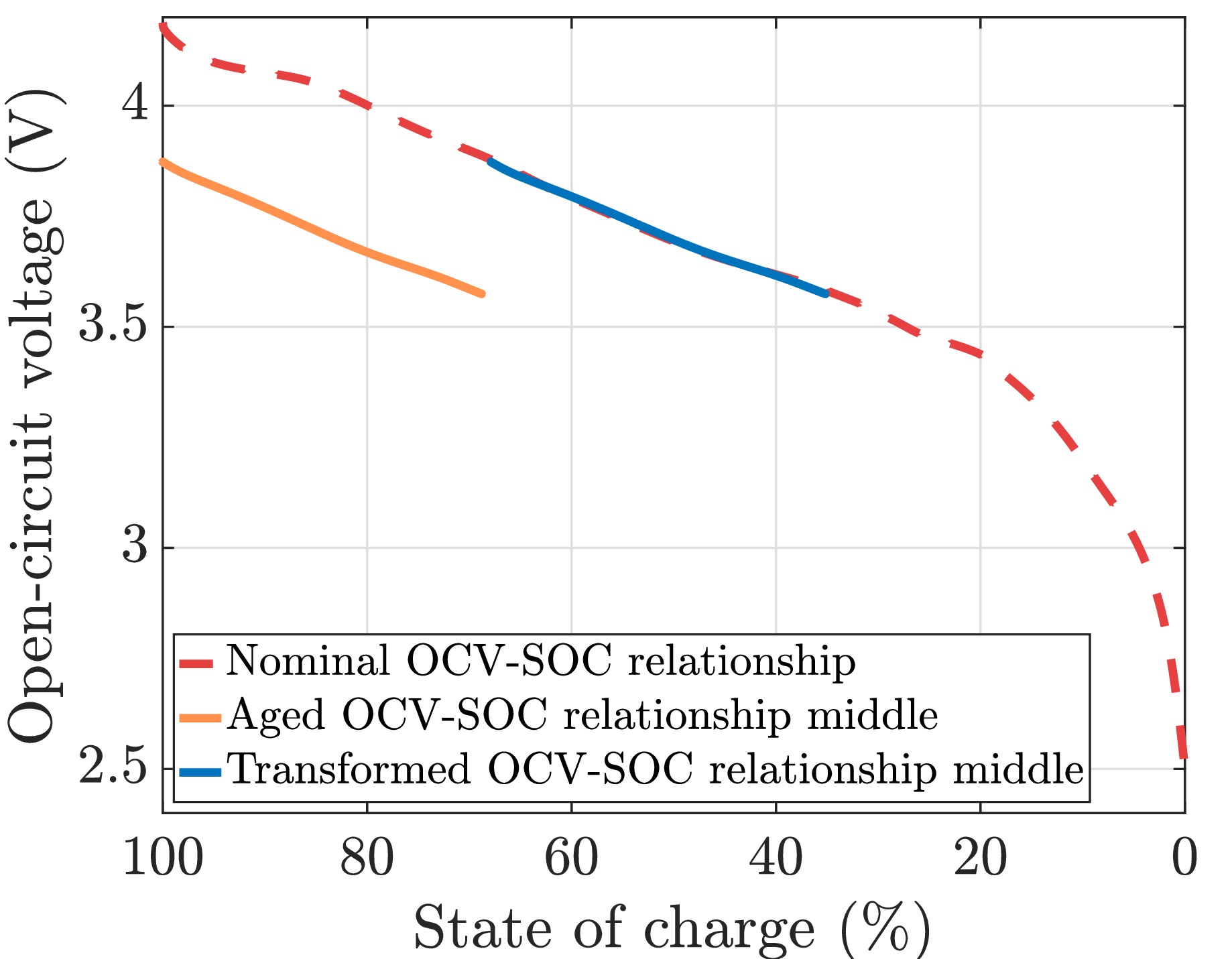}
        \caption{Cycle 194 middle}
        \label{fig:cycle194_middle}
    \end{subfigure}

    \vspace{\baselineskip}

    \begin{subfigure}[t]{0.24\textwidth}
        \centering
        \includegraphics[width=\textwidth]{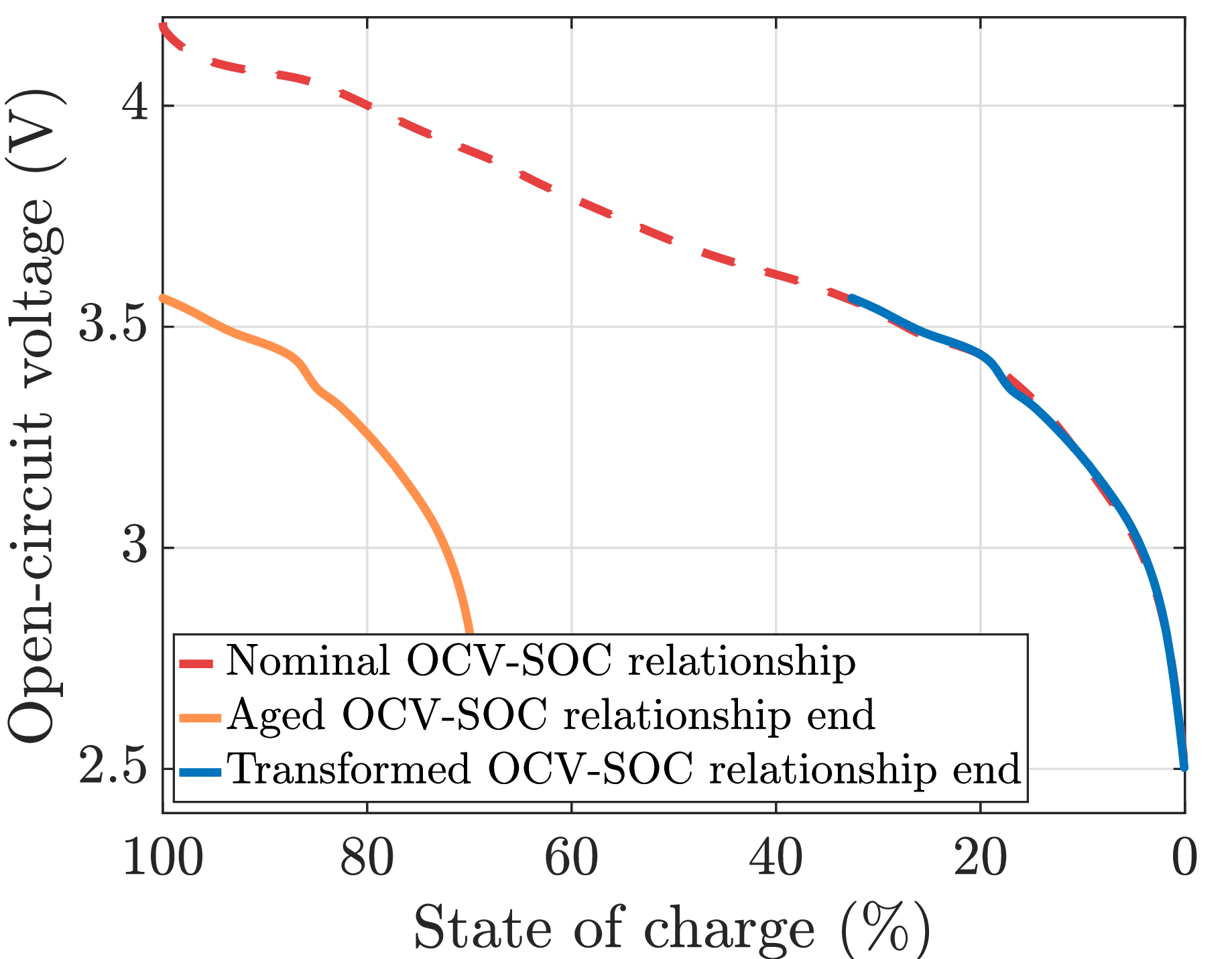}
        \caption{Cycle 159 end}
        \label{fig:cycle159_end}
    \end{subfigure}
    \hfill
    \begin{subfigure}[t]{0.24\textwidth}
        \centering
        \includegraphics[width=\textwidth]{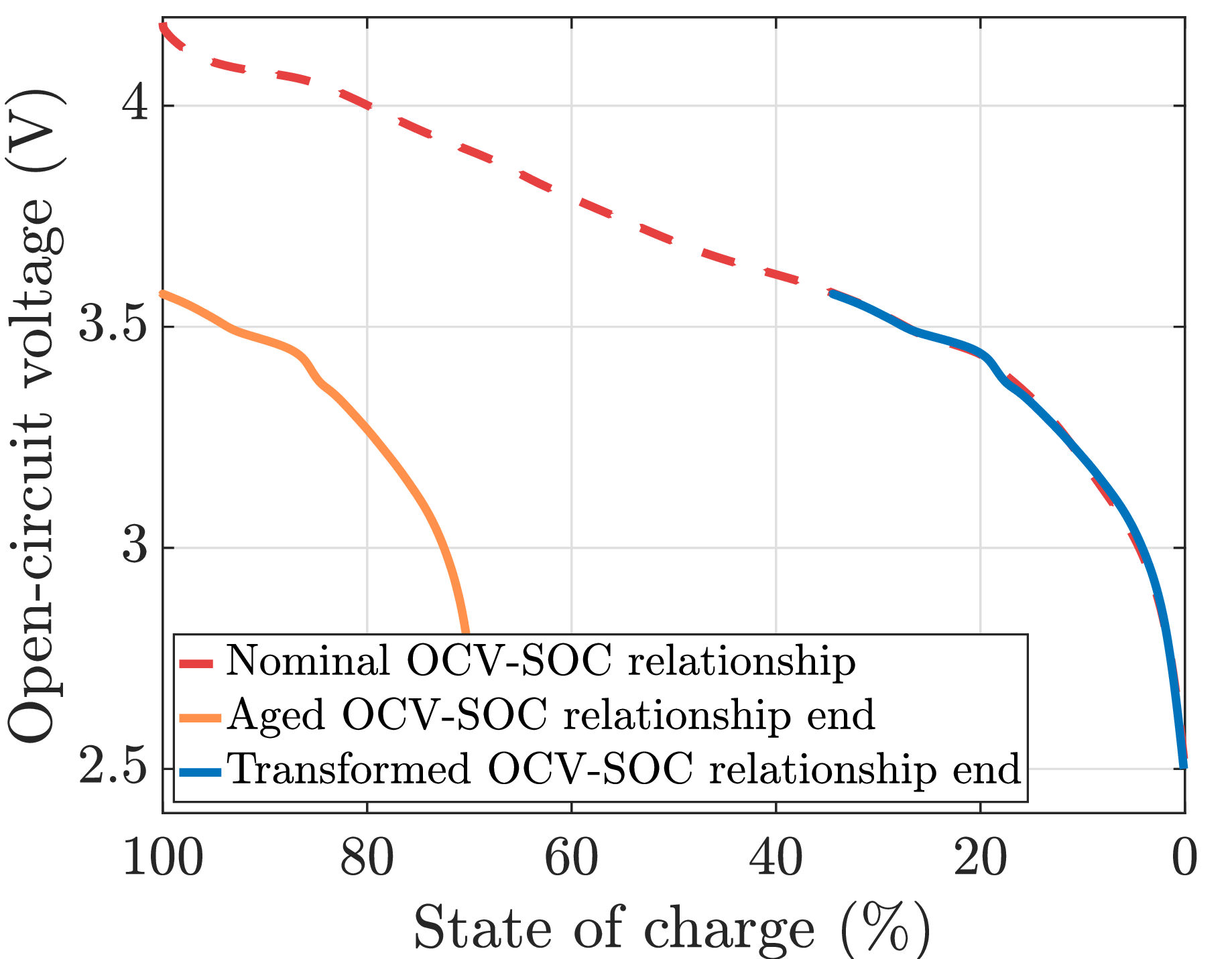}
        \caption{Cycle 194 end}
        \label{fig:cycle194_end}
    \end{subfigure}

    \caption{OCV-SOC transformations of capacity estimation using partial data of the OCV tests. (a)(c)(e): Transformation for the 159th cycle. (b)(d)(f): Transformation for the 194th cycle.}
    \label{fig:capacity_estimation_OCVtest_partial}
\end{figure}

The OCV-SOC transformations of two cycles using each part of the OCV data are shown in Figure \ref{fig:capacity_estimation_OCVtest_partial}. From the figures, we see that the transformations achieve good alignment to the nominal relationships with partial OCV data from the beginning, middle, and end parts of the cycles. 

This result demonstrates the effectiveness of the developed method on capacity estimation using the invariance of the OCV-SOC relationship of the aged battery.

\subsection{Capacity estimation from dynamic discharge data}
We estimate the capacity using the OCV values identified from dynamic discharge data with the OCV identification algorithm \citep{wang2025directcontinuoustimelpvidentification}. The OCV data identified from the 1st and the 344th cycles are used to validate the capacity estimation. 
The identification RMSEs of the terminal voltage for both cycles are 4.3 mV and 4.0 mV, respectively, 0.1\% of the upper cut-off voltage. 

The estimated capacities, actual capacities, and the AREs of the estimation are tabulated in Table \ref{tab: capacity estimation identified OCV}.

\begin{table}[htbp!]
\centering
\caption{Estimated capacities (\%) and absolute relative errors (\%) of capacity estimation using identified OCV relationships.}
\begin{tabular}{llll}
\hline
Cycle & \begin{tabular}[c]{@{}l@{}}Estimated \\ capacity (Ah)\end{tabular} & \begin{tabular}[c]{@{}l@{}}Actual \\ capacity (Ah)\end{tabular} & ARE (\%) \\ \hline
Cycle 1     & 4.8985                                                             & 4.8566                                                          & 0.8640   \\
Cycle 344   & 4.4479                                                             & 4.4505                                                          & 0.0604   \\ \hline
\end{tabular}
\label{tab: capacity estimation identified OCV}
\end{table}

We see from the table that with the OCV identified from dynamic data, the capacity estimation is accurate for both the first cycle and the last cycle of the aging test with AREs of less than 0.9\%. These results demonstrate the efficacy of combining an OCV identification algorithm with the developed capacity estimation.

\begin{figure}[htbp!]
    \centering
    \begin{subfigure}[t]{0.24\textwidth}
        \centering
        \includegraphics[width=\textwidth]{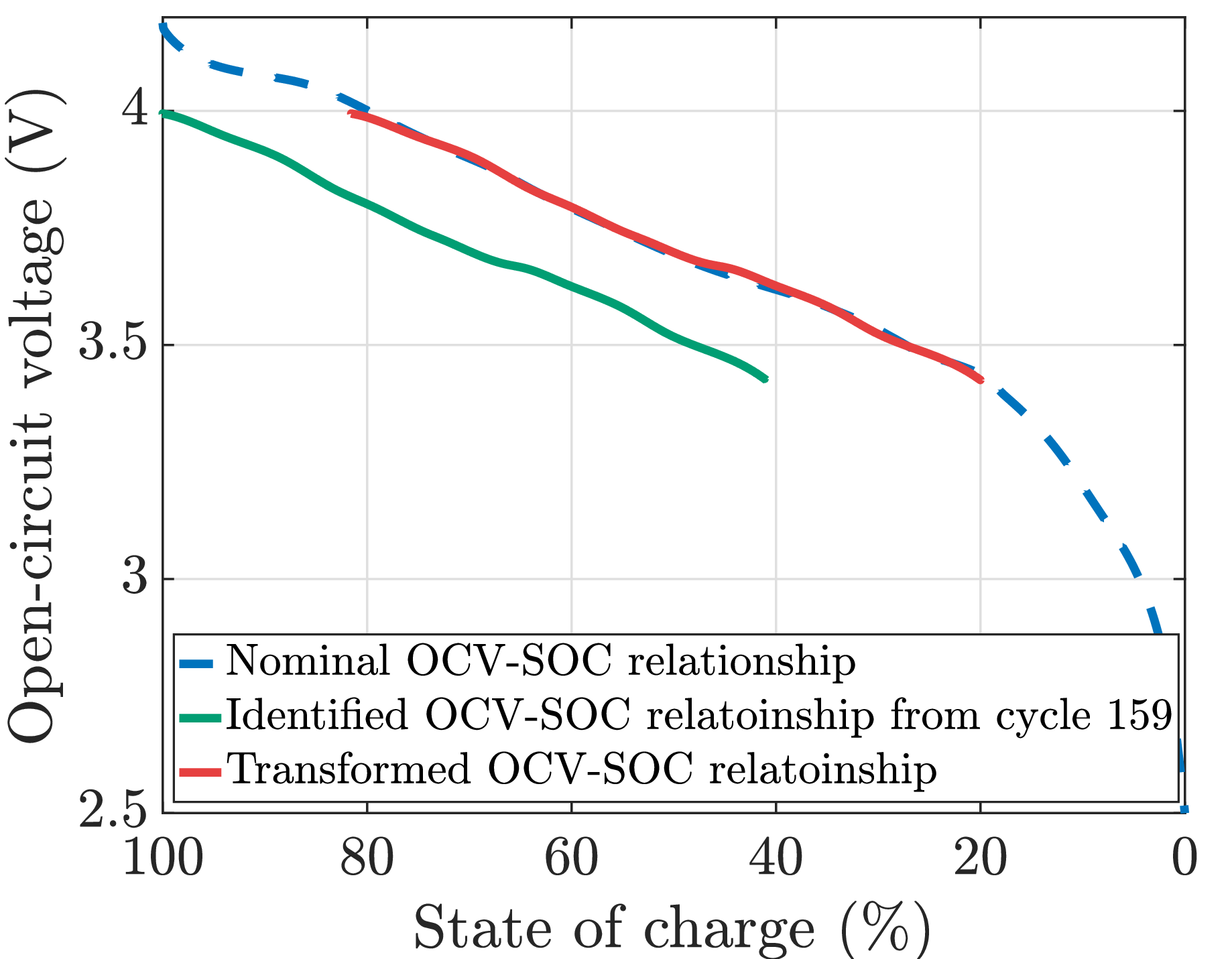}
        \caption{Transformation of the identified OCV-SOC relationship from cycle 159.}
    \end{subfigure}
    \hfill
    \begin{subfigure}[t]{0.24\textwidth}
        \centering
        \includegraphics[width=\textwidth]{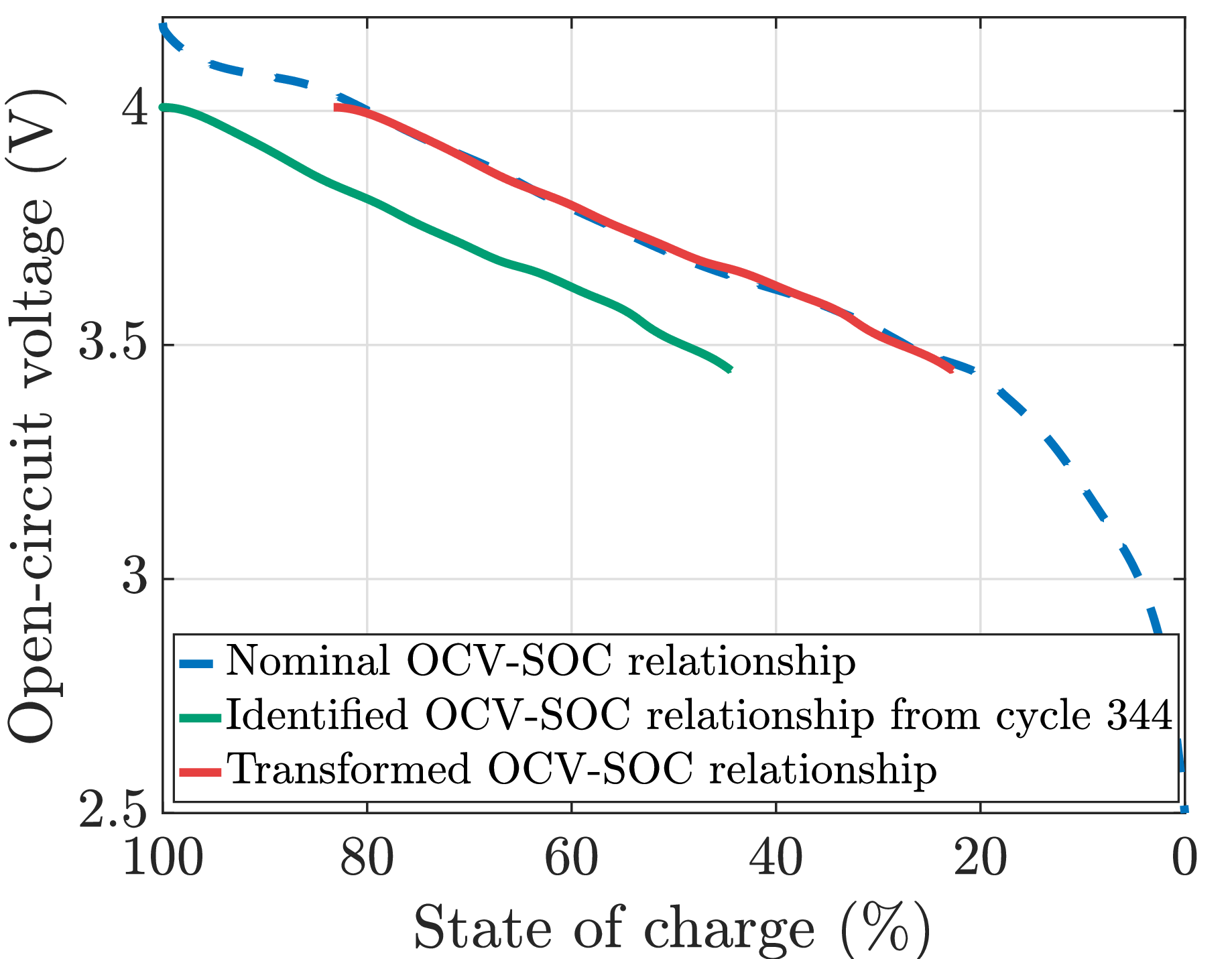}
        \caption{Transformation of the identified OCV-SOC relationship from cycle 344.}
    \end{subfigure}
        
    \caption{Transformations of the identified OCV-SOC relationships of capacity estimation from cycles 159 and 344.}
    \label{fig: OCV-Q curve comparison}
\end{figure}

Figure \ref{fig: OCV-Q curve comparison} shows the OCV-SOC transformation with the estimated capacities of the two cycles in comparison to the nominal relationship from the first cycle. We see from the figure that the transformed OCV-SOC relationships align with the nominal relationship with satisfactory accuracy.

\begin{figure}[htbp!]
    \centering
    \includegraphics[width=1\linewidth]{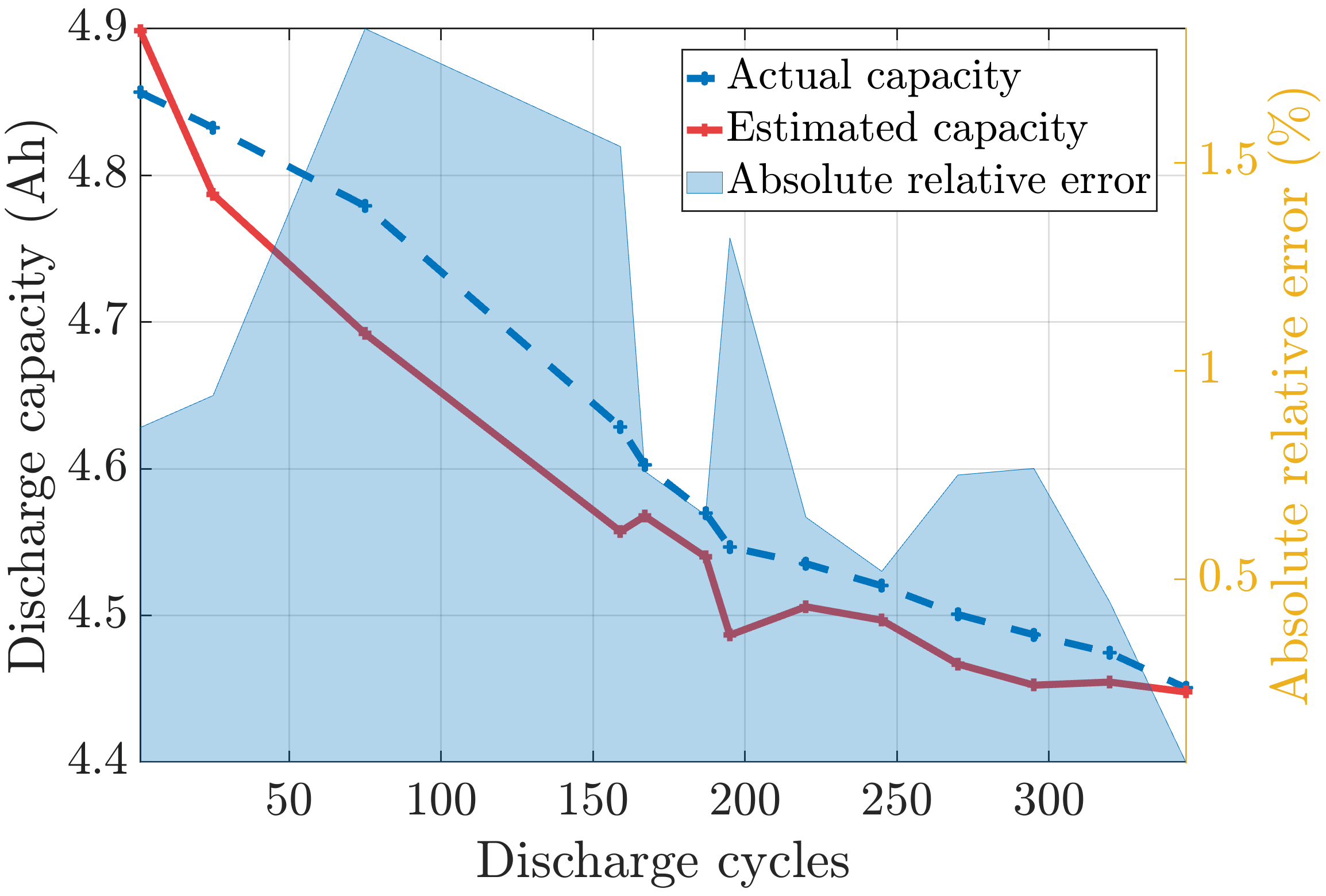}
    \caption{Estimated capacity in comparison to the actual capacity of the aged battery after dynamic cycles.}
    \label{fig:estimated capacity comparison}
\end{figure}

We adopted the developed approach to 12 test cycles from the 344 aging cycles using the OCVs identified from dynamic discharge data. The estimated capacity in comparison with the actual capacity and the ARE of the estimation are shown in Figure \ref{fig:estimated capacity comparison}. From the figure, we see that the developed method is effective in capturing the decreasing trend in the capacity and estimates the capacity with a satisfactory accuracy of less than 2\% ARE.

The average RMSE and MAE of the capacity estimation for all cycles are 0.045 Ah and 0.040 Ah, respectively. The mean ARE of the estimation is 0.85\%. The accuracy across aging cycles demonstrates the efficacy of the developed method for estimating the capacity of the aged battery with OCV-SOC relationships identified from dynamic discharge data.

\section{Discussion}
The developed approach principally relies on the invariance assumption on the OCV-SOC relationship. Although it is empirically validated for the studied battery, it is recommended to conduct an experiment to validate the invariance for the batteries that differ from the studied one. Additionally, the temperature of battery operation is assumed to be constant in this work. When the ambient temperature in real operation differs from the experimental condition, an additional test for invariance may be needed. According to the simulation results, the middle part of the OCV curve likely provides a higher estimation accuracy of the capacity. For the length of the estimation data, it is recommended to have a longer horizon to enhance estimation accuracy.

\section{Conclusion}\label{sec: conclusion}

In this work, we developed a novel method that estimates a battery's capacity based on its discharge capacity and the nominal OCV-SOC relationship. The estimation is achieved through an invariance property in the OCV-SOC relationship across aging cycles. With the invariance, the capacity can be estimated by solving an OCV alignment optimization problem using the OCV of the aged battery. This approach applies to partial OCV data without using a full cycle of battery operation. With an OCV identification algorithm, the capacity can be estimated from dynamic discharge data. Simulation experiments showed an average ARE of capacity estimation with OCV test data of less than 0.5\% and an average ARE of 0.85\% using dynamic discharge data across 12 test cycles from 344 aging cycles, sufficient for practical usage in EV batteries. In future work, we will extend the developed approach to predict the remaining useful life of the battery.



\bibliography{ifacconf}             

@article{schmitt2023capacity,
  title={Capacity and degradation mode estimation for lithium-ion batteries based on partial charging curves at different current rates},
  author={Schmitt, Julius and Rehm, Mathias and Karger, Alexander and Jossen, Andreas},
  journal={Journal of Energy Storage},
  volume={59},
  pages={106517},
  year={2023},
  publisher={Elsevier}
}

@article{dubarry2012synthesize,
  title={Synthesize battery degradation modes via a diagnostic and prognostic model},
  author={Dubarry, Matthieu and Truchot, Cyril and Liaw, Bor Yann},
  journal={Journal of power sources},
  volume={219},
  pages={204--216},
  year={2012},
  publisher={Elsevier}
}

@article{zagorowska2020survey,
  title={A survey of models of degradation for control applications},
  author={Zagorowska, Marta and Wu, Ouyang and Ottewill, James R and Reble, Marcus and Thornhill, Nina F},
  journal={Annual Reviews in Control},
  volume={50},
  pages={150--173},
  year={2020},
  publisher={Elsevier}
}

@misc{wang2025continuoustimeidentificationocvreconstruction,
      title={Continuous-Time System Identification and OCV Reconstruction of Li-ion Batteries via Regularized Least Squares}, 
      author={Yang Wang and Riccardo M. G. Ferrari and Michel Verhaegen},
      year={2025},
      eprint={2509.21116},
      archivePrefix={arXiv},
      primaryClass={eess.SY},
      url={https://arxiv.org/abs/2509.21116}, 
}

@misc{wang2025directcontinuoustimelpvidentification,
      title={Direct Continuous-Time LPV System Identification of Li-ion Batteries via L1-Regularized Least Squares}, 
      author={Yang Wang and Riccardo M. G. Ferrari},
      year={2025},
      eprint={2509.21110},
      archivePrefix={arXiv},
      primaryClass={eess.SY},
      url={https://arxiv.org/abs/2509.21110}, 
}

@article{lianpo2025capacity,
  title={Capacity degradation prediction of electric vehicle battery by integrating convolutional neural network with informer model},
  author={Lianpo, Li and Songmei, Dong and Lin, Wang},
  journal={Journal of Power Sources},
  volume={651},
  pages={237497},
  year={2025},
  publisher={Elsevier}
}

@article{lu2022battery,
  title={Battery degradation prediction against uncertain future conditions with recurrent neural network enabled deep learning},
  author={Lu, Jiahuan and Xiong, Rui and Tian, Jinpeng and Wang, Chenxu and Hsu, Chia-Wei and Tsou, Nien-Ti and Sun, Fengchun and Li, Ju},
  journal={Energy Storage Materials},
  volume={50},
  pages={139--151},
  year={2022},
  publisher={Elsevier}
}

@article{li2024online,
  title={An online state-of-health estimation method for lithium-ion battery based on linear parameter-varying modeling framework},
  author={Li, Yong and Wang, Liye and Feng, Yanbiao and Liao, Chenglin and Yang, Jue},
  journal={Energy},
  volume={298},
  pages={131277},
  year={2024},
  publisher={Elsevier}
}

@article{el2023physics,
  title={Physics-based model informed smooth particle filter for remaining useful life prediction of lithium-ion battery},
  author={El-Dalahmeh, Mo'ath and Al-Greer, Maher and El-Dalahmeh, Ma'd and Bashir, Imran},
  journal={Measurement},
  volume={214},
  pages={112838},
  year={2023},
  publisher={Elsevier}
}

@article{bloom2005differential,
  title={Differential voltage analyses of high-power, lithium-ion cells: 1. Technique and application},
  author={Bloom, Ira and Jansen, Andrew N and Abraham, Daniel P and Knuth, Jamie and Jones, Scott A and Battaglia, Vincent S and Henriksen, Gary L},
  journal={Journal of Power Sources},
  volume={139},
  number={1-2},
  pages={295--303},
  year={2005},
  publisher={Elsevier}
}

@article{navidi2024physics,
  title={Physics-informed machine learning for battery degradation diagnostics: A comparison of state-of-the-art methods},
  author={Navidi, Sina and Thelen, Adam and Li, Tingkai and Hu, Chao},
  journal={Energy Storage Materials},
  volume={68},
  pages={103343},
  year={2024},
  publisher={Elsevier}
}

@article{wang2023review,
  title={A review on rapid state of health estimation of lithium-ion batteries in electric vehicles},
  author={Wang, Zuolu and Zhao, Xiaoyu and Fu, Lei and Zhen, Dong and Gu, Fengshou and Ball, Andrew D},
  journal={Sustainable Energy Technologies and Assessments},
  volume={60},
  pages={103457},
  year={2023},
  publisher={Elsevier}
}

@article{peng2025state,
  title={State of health estimation joint improved grey wolf optimization algorithm and LSTM using partial discharging health features for lithium-ion batteries},
  author={Peng, Simin and Wang, Yujian and Tang, Aihua and Jiang, Yuxia and Kan, Jiarong and Pecht, Michael},
  journal={Energy},
  volume={315},
  pages={134293},
  year={2025},
  publisher={Elsevier}
}

@article{zhang2025study,
  title={Study on battery capacity recognition method for real electric vehicle under complex operating conditions},
  author={Zhang, Kai and Lv, Taolin and Chen, Xingguang and Chen, Jianguo and Shen, Yifan and Zhou, Long and Sun, Tao and Zheng, Yuejiu},
  journal={Journal of Energy Storage},
  volume={118},
  pages={116219},
  year={2025},
  publisher={Elsevier}
}

@article{wan2025degradation,
  title={Degradation and expansion of lithium-ion batteries with silicon/graphite anodes: Impact of pretension, temperature, C-rate and state-of-charge window},
  author={Wan, Zhiwen and Pannala, Sravan and Solbrig, Charles and Garrick, Taylor R and Stefanopoulou, Anna G and Siegel, Jason B},
  journal={eTransportation},
  volume={24},
  pages={100416},
  year={2025},
  publisher={Elsevier}
}

@book{plett2015battery,
  title={Battery management systems, Volume I: Battery modeling},
  author={Plett, Gregory L},
  volume={1},
  year={2015},
  publisher={Artech House}
}

@article{hannan2017review,
  title={A review of lithium-ion battery state of charge estimation and management system in electric vehicle applications: Challenges and recommendations},
  author={Hannan, Mohammad A and Lipu, MS Hossain and Hussain, Aini and Mohamed, Azah},
  journal={Renewable and Sustainable Energy Reviews},
  volume={78},
  pages={834--854},
  year={2017},
  publisher={Elsevier}
}

@article{meng2018overview,
  title={Overview of lithium-ion battery modeling methods for state-of-charge estimation in electrical vehicles},
  author={Meng, Jinhao and Luo, Guangzhao and Ricco, Mattia and Swierczynski, Maciej and Stroe, Daniel-Ioan and Teodorescu, Remus},
  journal={Applied sciences},
  volume={8},
  number={5},
  pages={659},
  year={2018},
  publisher={MDPI}
}

@article{mcturk2015minimally,
  title={Minimally invasive insertion of reference electrodes into commercial lithium-ion pouch cells},
  author={McTurk, E and Birkl, CR and Roberts, MR and Howey, DA and Bruce, PG},
  journal={ECS Electrochemistry Letters},
  volume={4},
  number={12},
  pages={A145},
  year={2015},
  publisher={IOP Publishing}
}

@article{pozzato2022lithium,
  title={Lithium-ion battery aging dataset based on electric vehicle real-driving profiles},
  author={Pozzato, Gabriele and Allam, Anirudh and Onori, Simona},
  journal={Data in brief},
  volume={41},
  pages={107995},
  year={2022},
  publisher={Elsevier}
}
                                                   







\end{document}